\PassOptionsToPackage{dvipsnames}{xcolor}
\documentclass[letterpaper]{article}
\usepackage{authblk}
\usepackage{endnotes}
\usepackage{epsfig}
\usepackage{graphicx}
\usepackage{dcolumn}
\setkeys{Gin}{width=2in}
\usepackage{xspace}
\PassOptionsToPackage{hyphens}{url}
\usepackage{url}
\usepackage[squaren, binary]{SIunits}
\usepackage{amsmath}
\usepackage{paralist}
\usepackage{multirow}
\usepackage{rotating}
\usepackage{subcaption}
\usepackage{balance}
\usepackage{comment}
\usepackage{etoolbox}
\usepackage{booktabs}
\usepackage{xcolor}
\usepackage{algorithm}
\usepackage[noend]{algpseudocode}
\usepackage{environ}
\usepackage{xargs}
\usepackage{csquotes}
\usepackage{tikz}
\usepackage{pgfplots}
\usepackage{pgfplotstable}
\usepackage{pgfkeys}
\usepackage{pgffor}
\usepgfplotslibrary{external}
\pgfplotsset{compat=1.17,
   invoke before crossref tikzpicture={\tikzexternaldisable},
   invoke after crossref tikzpicture={\tikzexternalenable}
}
\usepackage{fp}
\usepackage{wrapfig}
\usepackage{enumitem}
\usepackage{hyperref}
\usepackage[noabbrev]{cleveref}
\usepackage[overload]{bibalias}

\renewcommand{\epsilon}{\varepsilon}

\newcommand{\bet}{B$^{\varepsilon}$-tree\xspace}
\newcommand{\bets}{B$^{\varepsilon}$-trees\xspace}

\newcommand{\io}{IO\xspace}
\newcommand{\ios}{IOs\xspace}

\newcommand{\ext}{{ext4}\xspace}
\newcommand{\Extt}{{Ext4}\xspace}

\newcommand{\xfs}{{XFS}\xspace}
\newcommand{\zfs}{{ZFS}\xspace}
\newcommand{\ftwofs}{{F2FS}\xspace}

\newcommand{\defn}[1]{\textit{\textbf{\boldmath #1}}}

\setdefaultleftmargin{1em}{0em}{}{}{}{}
\newcolumntype{d}[1]{D{.}{.}{#1}}

\definecolor{cerisepink}{rgb}{0.93, 0.23, 0.51}
\definecolor{ballblue}{rgb}{0.13, 0.67, 0.8}

\newcommand{\punt}[1]{}

\newcommand{\betrfs}{BetrFS\xspace}

\newcommand{\btrfs}{{Btrfs}\xspace}
\newcommand{\Btrfs}{{Btrfs}\xspace}

\newcommand{\blktrace}{\texttt{blktrace}\xspace}
\newcommand{\efreefrag}{\texttt{e2freefrag}\xspace}
\newcommand{\proc}[1]		{\ifmmode\mbox{\textsc{#1}}\else\textsc{#1}\fi}

\newcommand{\fsync}{\texttt{fsync}\xspace}

\newcommand{\linuxver}{{$3.11.10$}\xspace}
\newcommand{\secput}[2]{\section{#2}\label{sec:#1}}
\newcommand{\secref}[1]{Section~\ref{sec:#1}}
\newcommand{\subsecput}[2]{\subsection{#2}\label{sec:#1}}

\newcommand{\figref}[1]{Figure~\ref{fig:#1}}

\title{File System Aging}

  \author[1,2]{Alex Conway}
  \author[3]{Ainesh Bakshi}
  \author[4]{Arghya Bhattacharya}
  \author[4]{Rory Bennett}
  \author[5]{Yizheng Jiao}
  \author[6]{Eric Knorr}
  \author[7]{Yang Zhan}
  \author[4]{Michael A. Bender}
  \author[8]{William Jannen}
  \author[2]{Rob Johnson}
  \author[9]{Bradley C. Kuszmaul}
  \author[5]{Donald E. Porter}
  \author[10]{Jun Yuan}
  \author[11]{Martin Farach-Colton}
  \affil[1]{Cornell Tech}
  \affil[2]{VMware Research}
  \affil[3]{Carnegie Mellon University}
  \affil[4]{Stony Brook University}
  \affil[5]{University of North Carolina at Chapel Hill}
  \affil[6]{Harvard University}
  \affil[7]{Huawei}
  \affil[8]{Williams College}
  \affil[9]{Google}
  \affil[10]{Data Dog}
  \affil[11]{Rutgers University}

\begin{document}

\maketitle

\begin{abstract}

   File systems must allocate space for files without knowing what will be
   added or removed in the future. Over the life of a file system, this may
   cause suboptimal file placement decisions that eventually lead to slower
   performance, or \defn{aging}. Traditional file systems employ heuristics,
   such as colocating related files and data blocks, to avoid aging, and many
   file system implementors treat aging as a solved problem in the common case,
   but it is believed that when a storage device fills up, space pressure
   exacerbates fragmentation-based aging.
	
   However, this article describes both realistic and synthetic workloads that
   can cause these heuristics to fail, inducing large performance declines due
   to aging even when the storage device is nearly empty. For example, on \ext
   and \zfs, a few thousand git pull operations can reduce read performance by a
   factor of 2, and performing 10000 pulls can reduce performance by up to
   a factor of 8. 
   We further present microbenchmarks demonstrating that
   common placement strategies are extremely sensitive to file-creation order;
   varying the creation order of a few thousand small files in a real-world
   directory structure can slow down reads by 2--10$\times$ on hard disks, 
   depending on the file system.

   We argue that these slowdowns are caused by poor layout. We demonstrate a
   correlation between the read performance of a directory scan and the locality
   within a file system's access patterns, using a \defn{dynamic layout score}.

   We complement these results with microbenchmarks that show that space
   pressure can cause a substantial amount of inter-file and intra-file
   fragmentation.  However, on a ``real-world'' application benchmark, space
   pressure causes fragmentation that slows subsequent reads by only 20\% on
   \ext, relative to the amount of fragmentation that would occur on a file
   system with abundant space.  The other file systems show negligible
   additional degradation under space pressure.

   Our results suggest that the effect of free-space fragmentation on read
   performance is best described as accelerating the file system aging process.
   The effect on write performance is non-existent in some cases, and, in most
   cases, an order of magnitude smaller than the read degradation from
   fragmentation caused by normal usage.
	
   In short, many file systems are exquisitely prone to read aging after a
   variety of write patterns. We show, however, that aging is not inevitable.
   \betrfs, a file system based on write-optimized dictionaries, exhibits
   almost no aging in our experiments.  \betrfs typically outperforms the other
   file systems in our benchmarks; aged \betrfs even outperforms the unaged
   versions of these file systems, excepting \btrfs.  We present a framework
   for understanding and predicting aging, and identify the key features of
   \betrfs that avoid aging.

\end{abstract}

\pgfkeys{
	/fs-names/ext4/.initial=\ext,
	/fs-names/btrfs/.initial=\btrfs,
	/fs-names/xfs/.initial=\xfs,
	/fs-names/f2fs/.initial=\ftwofs
}
\pgfkeys{
	/fs-colors/ext4/.initial=blue,
	/fs-colors/btrfs/.initial=red,
	/fs-colors/xfs/.initial=green!60!black,
	/fs-colors/f2fs/.initial=orange
}
\pgfkeys{
	/fs-fullness-marks/ext4-full/.initial=*,
	/fs-fullness-marks/btrfs-full/.initial=square*,
	/fs-fullness-marks/xfs-full/.initial=diamond*,
	/fs-fullness-marks/f2fs-full/.initial=triangle*,
	/fs-fullness-marks/ext4-95/.initial=*,
	/fs-fullness-marks/btrfs-95/.initial=square*,
	/fs-fullness-marks/xfs-95/.initial=diamond*,
	/fs-fullness-marks/f2fs-95/.initial=triangle*,
	/fs-fullness-marks/ext4-90/.initial=*,
	/fs-fullness-marks/btrfs-90/.initial=square*,
	/fs-fullness-marks/xfs-90/.initial=diamond*,
	/fs-fullness-marks/f2fs-90/.initial=triangle*,
	/fs-fullness-marks/ext4-empty/.initial=o,
	/fs-fullness-marks/btrfs-empty/.initial=square,
	/fs-fullness-marks/xfs-empty/.initial=diamond,
	/fs-fullness-marks/f2fs-empty/.initial=triangle,
	/fs-fullness-marks/ext4-10/.initial=o,
	/fs-fullness-marks/btrfs-10/.initial=square,
	/fs-fullness-marks/xfs-10/.initial=diamond,
	/fs-fullness-marks/f2fs-10/.initial=triangle,
	/fs-fullness-marks/ext4-clean/.initial=o,
	/fs-fullness-marks/btrfs-clean/.initial=square,
	/fs-fullness-marks/xfs-clean/.initial=diamond,
	/fs-fullness-marks/f2fs-clean/.initial=triangle,
}
\pgfkeys{
	/fullness-dashes/full/.initial=solid,
	/fullness-dashes/empty/.initial=dashed,
	/fullness-dashes/95/.initial=solid,
	/fullness-dashes/90/.initial=solid,
	/fullness-dashes/10/.initial=dashed,
	/fullness-dashes/clean/.initial=dotted
}
\pgfkeys{
	/fullness-mark-options/full/.initial=solid,
	/fullness-mark-options/empty/.initial=solid,
	/fullness-mark-options/95/.initial=solid,
	/fullness-mark-options/90/.initial=solid,
	/fullness-mark-options/10/.initial=solid,
	/fullness-mark-options/clean/.initial=dashed
}
\newcommand{\plotlinewidth}{0.5pt}

\newcommand{\fses}{betrfs, btrfs, ext4, f2fs, xfs, zfs}
\newcommand{\agednesses}{clean, aged}
\newcommand{\hardwares}{ssd, hdd, ssd_raoff}
\newcommand{\partitionsizes}{4gb, 20gb}

\pgfkeys{
  /fs-names/betrfs/.initial=\betrfs,
  /fs-names/btrfs/.initial=\btrfs,
  /fs-names/ext4/.initial=\ext,
  /fs-names/f2fs/.initial=\ftwofs,
  /fs-names/xfs/.initial=\xfs,
  /fs-names/zfs/.initial=\zfs
}

\pgfkeys{
  /agedness-names/clean/.initial=clean,
  /agedness-names/aged/.initial=aged
}

\pgfkeys{
  /hardware-names/ssd/.initial=SSD,
  /hardware-names/hdd/.initial=HDD,
  /hardware-names/ssd_raoff/.initial={SSD, no read-ahead}
}

\pgfkeys{
  /git-gc-mode-names/on/.initial=on,
  /git-gc-mode-names/off/.initial=off
}

\pgfkeys{
  /partition-size-names/4gb/.initial={4GB},
  /partition-size-names/20gb/.initial={20GB}
}

\pgfkeys{
  /fs-colors/betrfs/.initial=Black,
  /fs-colors/btrfs/.initial=Red,
  /fs-colors/ext4/.initial=Plum,
  /fs-colors/f2fs/.initial=Aquamarine,
  /fs-colors/xfs/.initial=LimeGreen,
  /fs-colors/zfs/.initial=Dandelion
}

\pgfkeys{
  /fs-marks/betrfs/.initial=*,
  /fs-marks/btrfs/.initial=pentagon*,
  /fs-marks/ext4/.initial=triangle*,
  /fs-marks/f2fs/.initial=x,
  /fs-marks/xfs/.initial=square*,
  /fs-marks/zfs/.initial=diamond*
}

\pgfkeys{
  /agedness-styles/clean/.initial=dashed,
  /agedness-styles/aged/.initial=solid
}

\pgfkeys{
  /ram-sizes/768mb/.initial={768MiB},
  /ram-sizes/1024mb/.initial={1024MiB},
  /ram-sizes/1280mb/.initial={1280MiB},
  /ram-sizes/1536mb/.initial={1536MiB},
  /ram-sizes/2048mb/.initial={2048MiB},
  /ram-sizes/cold/.initial={Cold Cache Aged},
  /ram-sizes/cold_clean/.initial={Cold Cache Unaged},
  /ram-sizes/betrfs/.initial={\betrfs Cold Cache Aged},
}

\pgfkeys{
  /rs-marks/768mb/.initial=*,
  /rs-marks/1024mb/.initial=pentagon*,
  /rs-marks/1280mb/.initial=triangle*,
  /rs-marks/1536mb/.initial=square*,
  /rs-marks/2048mb/.initial=diamond*,
  /rs-marks/cold/.initial=otimes,
  /rs-marks/cold_clean/.initial=x,
  /rs-marks/betrfs/.initial=oplus,
}

\pgfkeys{
  /rs-colors/768mb/.initial=Plum,
  /rs-colors/1024mb/.initial=Red,
  /rs-colors/1280mb/.initial=LimeGreen,
  /rs-colors/1536mb/.initial=CarnationPink,
  /rs-colors/2048mb/.initial=Dandelion,
  /rs-colors/cold/.initial=RoyalBlue,
  /rs-colors/cold_clean/.initial=Cyan,
  /rs-colors/betrfs/.initial=Black,
}

\secput{intro}{Introduction}

File systems tend to slow over time, or \defn{age}, as they become increasingly fragmented as files are created,
deleted, moved, appended to, and truncated~\cite{SmithSe97, McKusickJoLe84}.

Fragmentation occurs when logically contiguous file blocks---either from a large file or from small files in the same directory---become scattered on disk.
Reading these files requires additional seeks, and on hard drives, a few seeks
can have an outsized effect on performance.  For example, if a file system
places a 100\mebi\byte{} file in 200 disjoint pieces (i.e.,
200 seeks) on a disk with 100\mebi\byte\per\second{} bandwidth
and 5\milli\second{} seek time, reading the data will take twice as long
as reading it in an ideal layout (i.e., one seek).
Even on SSDs, which do not perform mechanical seeks,
a decline in logical block locality can harm performance~\cite{MinKiCh12}.

The state of the art in mitigating file system \defn{aging} 
applies best-effort heuristics at allocation time to prevent fragmentation.  
For example, file systems attempt to
place related files close together on disk while also leaving empty space for
future files~\cite{McKusickJoLe84, CardTsTw94, Tweedie00, MathurCaBh07}.
In addition, some file systems (including \ext, \xfs, \btrfs, and \ftwofs, among those tested in
this article) attempt to reverse aging by including defragmentation tools that reorganize files and file blocks into contiguous regions.

Over the past two decades, there have been differing opinions about the
significance of aging.  The seminal work of Smith and Seltzer~\cite{SmithSe97}
showed that file systems age under realistic workloads, and this aging affects
performance.  On the other hand, there is a widely held view in the developer
community that aging is a solved problem in production file systems unless,
perhaps, the device is nearly full.  For example, the Linux System
Administrator's Guide~\cite{WirzeniusOjSt04} says:
\begin{displayquote}
  Modern Linux file systems keep fragmentation at a minimum by keeping
  all blocks in a file close together, even if they can't be stored in
  consecutive sectors. Some file systems, like ext3, effectively
  allocate the free block that is nearest to other blocks in a
  file. Therefore it is not necessary to worry about fragmentation in
  a Linux system.
\end{displayquote}

There have also been changes in storage technology and file-system design that
could substantially affect aging.  For example, a back-of-the-envelope analysis
suggests that aging should get worse as rotating disks get bigger, as seek
times have been relatively stable, but bandwidth grows (approximately) as the
square root of the capacity. Consider the same level of fragmentation as the
above example, but on a new, faster disk with 600\mebi\byte\per\second{} bandwidth but still a
5\milli\second{} seek time.  Then the 200 seeks would introduce a four-fold slowdown rather
than a two-fold slowdown. Thus, we expect fragmentation to become an
increasingly significant problem as the gap between random \io and sequential
\io grows.

As for SSDs, there is a widespread belief that fragmentation is not an issue.
For example, PCWorld measured the performance gains from defragmenting an NTFS
file system\cite{pcworld-ssd-defrag-benchmarks}, and concluded that,
``From my limited tests, I'm firmly convinced that the tiny difference that
even the best SSD defragger makes is not worth reducing the life span of your
SSD.''
Furthermore, SSD performance is more nuanced	
as SSDs have additional storage for over-provisioning, which helps
to improve SSD performance and prolong SSD lifetime.

In this article, we revisit the issue of file-system aging in light of changes
in storage hardware, file-system design, and data-structure theory.  We make
several contributions:
\begin{enumerate}
   \item We give a simple, fast, and portable method for aging file systems.

   \item We show that fragmentation over time (i.e., aging) is a first-order
      performance concern, and that this is true even on modern hardware, such
      as SSDs, even on modern file systems, and even when the storage device is
      nearly empty.

   \item We demonstrate a synthetic benchmark designed to stress the
      worst-case full-disk behavior of the file system. We show that although this
      benchmark can create more substantial aging on full disks than when there is no space pressure, the effect is modest on SSDs and substantially lower on HDDs, even on HDDs facing space pressure under some ``real-world'' aging benchmarks.

   \item Furthermore, we show that aging is not inevitable.
      We present several techniques for avoiding aging.
      We show that \betrfs~\cite{EsmetBeFa12,jannen15betrfs,JannenYuZh15tos,YuanZhJa16,yuan17tos,zhan18fast,zhan18tos,zhan20fast,zhan21tos},
      a research prototype that includes several of these design techniques, is
      much more resistant to aging than the other file systems we tested, at
      least when the device is not full.  In fact, \betrfs essentially did not
      age in our experiments on non-full disks, establishing that,
      aging is a solvable problem when disks are not full. In the near-full-disk setting, \betrfs was
      unable to complete the test suite because it became unstable. 
      It remains an open question whether \betrfs or other file systems can avoid
      aging under near-full-disk conditions.
\end{enumerate}

\paragraph{Results.} We use realistic application workloads to age---or degrade performance by inducing fragmentation---five
widely-used file systems---\btrfs~\cite{RodehBaMa13}, \ext~\cite{CardTsTw94,
Tweedie00, MathurCaBh07}, \ftwofs~\cite{lee15f2fs}, \xfs~\cite{SweeneyDoHu96},
and \zfs~\cite{BonwickMo08}---as well as the \betrfs research file system.  
One workload ages the file system by performing successive git checkouts of the
Linux kernel's source code repository, 
emulating the aging that a developer might experience on
their workstation.  A second workload ages the file system by running a
mail-server benchmark, emulating aging over the continued use of a server.

We evaluate the impact of aging as follows.  We periodically stop the aging
workload and measure the overall read throughput of the file system---more significant
fragmentation will result in slower read throughput.  To isolate the impact of
aging, as opposed to performance degradation due to changes in, say, the
distribution of file sizes, we then copy the file system onto a fresh
partition, essentially producing a defragmented or ``unaged'' version of the
file system, and we perform the same read throughput measurement.
We treat the differences in
read throughput between the aged and unaged copies as the result of aging.
Note that, using this methodology,
we focus exclusively on the performance impacts that aging induces on read operations.

\medskip

Our application benchmarks show that:
\begin{itemize}
   \item All the production file systems age on both hard disk drives (HDDs) and SSDs. For
      example, under our git workload, we observe over 62$\times$ slowdowns on
      HDDs and 2--6$\times$ slowdowns on SSDs. Similarly, under our mail-server
      workload, we observe 3--10$\times$ slowdowns on HDDs due to aging.
   \item Aging can happen quickly. For example, \ext shows over a 2$\times$
      slowdown after 1200 git pulls; \btrfs and \zfs slow down similarly after 1300 and 1600 pulls, respectively.  
   \item \betrfs exhibits essentially no aging for a few thousand git pulls. 
      Other than \btrfs, \betrfs's aged performance is close to the 
      other file systems' unaged performance on almost all benchmarks. 
   \item The costs of aging can be staggering in concrete terms. For example, at
      the end of our git workload on an HDD, all five production file systems took
      over 4 minutes to sequentially scan through 5\gibi\byte{} of data. 
      \ftwofs took over 50 minutes and \zfs over 60 minutes;
      \betrfs, on the other hand, took less than a minute.
\end{itemize}

We also performed several microbenchmarks to tease out specific causes of aging, and
we found that performance in the production file systems was sensitive to numerous
factors:
\begin{itemize}
   \item If only 10\% of files are created out of order relative to the
      directory structure (and therefore relative to a depth-first search of
      the directory tree) on HDDs, only \btrfs achieves a scan throughput of 
      75\mebi\byte\per\second{}, whereas \ext, \ftwofs, \xfs, and \zfs 
      achieve a scan throughput of only 19--40\mebi\byte\per\second{}. 
      If the files are copied completely out of order, then of these, only \xfs 
      achieves 23\mebi\byte\per\second{}, whereas \ext, \ftwofs, and \zfs 
      have a throughput of 6--9\mebi\byte\per\second{}. 
      These slowdowns are not inevitable;
      \betrfs
      throughput is 143\mebi\byte\per\second{} when files are copied in order,
      and it maintains a throughput of roughly 140\mebi\byte\per\second{}
      when 10\% of files are copied out of order.
      Yet, when files are copied completely out of order,
      \betrfs performance degrades to 13\mebi\byte\per\second{}.

   \item If an application writes to a file in small chunks, then the file's
      blocks can end up scattered on disk, harming performance when reading the
      file back.
      For example, in a benchmark that performs one hundred rounds of
      small appends to one hundred files on an HDD,
      \xfs and \zfs realize 31--40$\times$ lower read throughput
      than the baseline---when all files were
      written sequentially, one whole file at a time.
      \ftwofs ages by a factor of 11.
      \Extt and \btrfs are more stable but eventually age by a factor of 1.5.
      \betrfs throughput remains stable at one-third of the disk's raw
      bandwidth throughout the test.
      
   \item Disk fullness can amplify the read-throughput degradation caused by aging workloads, 
   although the impact of disk fullness is more pronounced on HDDs than on SSDs.
   We find that, on an HDD,
      a synthetic fragmentation benchmark ages \ext
      far worse on a full disk than on a nearly empty one.  For the
      other file systems, having a full disk roughly doubles the read-throughput
      degradation.
      On SSDs, disk fullness has a modest effect on 
      the read throughput degradation caused by the synthetic benchmark
      (typically less than 20\%), except on \btrfs.
      Disk fullness amplifies the read-throughput degradation caused by a git-based application benchmark on \ext by 20\% compared to an initially empty HDD.
      However, disk fullness has a negligible impact on the read-throughput degradation induced by the same git-based benchmark on \btrfs and \xfs on HDD,
      as well as for all file systems on SSD.
\end{itemize}

\secput{seq}{A Framework for Aging}

Because block devices can more efficiently access nearby disk addresses,
the relative proximity of related blocks directly affects a file system's performance.
\defn{Fragmentation} occurs when logically related blocks become scattered.
We categorize fragmentation by block type and their relationships:
\begin{itemize}
	\item \defn{Intrafile fragmentation:} fragmentation among a single file's allocated blocks.
	\item \defn{Interfile fragmentation:} fragmentation among the allocated blocks of small files that are in the
		same directory.
	\item \defn{Free-space fragmentation:} fragmentation among unallocated disk blocks.
\end{itemize}

The first two types of fragmentation directly impact the read performance of
a file system and therefore induce \defn{read aging}. 
When reading logically sequential data,
fragmented blocks will incur non-sequential reads, 
which on most modern storage hardware are considerably slower than sequential reads.

The impacts of free-space fragmentation on file system performance are more nuanced,
and thus we consider free-space fragmentation separately.
First, free-space fragmentation can affect read performance, but its impacts are indirect and already captured in the first two types of fragmentation.
To see why, consider a set of unallocated (free) blocks.
For these free blocks to be fragmented, they must be interspersed with allocated blocks, 
and the immediate impact of such interspersion is captured in our measures of inter- and intra-file fragmentation.
However, free space is where newly allocated blocks are drawn from, 
so some amount of free space is necessary for file systems to maintain locality as files and directories grow in size.
Thus, without appropriate free space fragmentation, \emph{future} allocations will introduce inter- and intra-file fragmentation and therefore lead to read aging.
These nuances are discussed further in \secref{freespacefragfullness}.

The above discussion hints at an important distinction between the related notions of 
fragmentation and aging.
Although we can quantify fragmentation at any point in time,
aging is a dynamic process induced by a sequence of file system operations over time;
the degree to which fragmentation worsens as a file system evolves determines the degree to which a file system ages.
Hence, to understand a file system's aging profile, we must treat the aging process as a path---where every file system operation produces a new point
(a static file system state)
on that path.

Now that we've categorized fragmentation along these three axes, the rest of this section will provide a framework for quantifying the degree of aging that we observe.

\subsecput{nts}{Natural Transfer Size}

Our aging model is based on the observation that
storage device bandwidth is typically maximized when \ios are large;
that is, sequential \ios are faster than random \ios.
We abstract away from hardware particulars
by defining the \defn{natural transfer size} (NTS) to be the minimum amount of
sequential data that must be transferred per \io in order to obtain some fixed
fraction of maximum throughput, say 50\% or 90\%.
\ios that exceed a device's NTS achieve an even larger fraction of the device's maximum bandwidth.

\figref{alpha-plot} plots SSD and HDD read bandwidth as a function of \io size.
From each device's address space,
we sampled 1000 offsets uniformly at random
and then performed multiple rounds of sequential reads.
Each round performed 1000 fixed-size reads, ranging from 4\kibi\byte--512\mebi\byte.
We conclude that a reasonable NTS for both the SSDs and
HDDs we measured is 4\mebi\byte.

The cause of the performance gap between sequential-\io and random-\io is different for
different hardware. For HDDs, seek times offer a simple explanation. 
For SSDs, this gap is hard to explain conclusively without vendor support;
common theories include: sequential accesses are easier to stripe across
internal banks, better-leveraging parallelism~\cite{JuKa13}; some FTL
translation data structures have nonuniform search times~\cite{MaFeLi14}; and
fragmented SSDs are not able to prefetch data~\cite{ChenKoZh09} or
metadata~\cite{JiChSh16}. Whatever the reason, SSDs show a modest gap between
sequential and random read performance, though not as great as on disks.

In order to avoid read aging, file systems should avoid breaking large files
into pieces significantly smaller than the NTS of the hardware. They should
also group small files that are logically related
(therefore likely to be accessed together)
into clusters of size at least the NTS and store the clusters at nearby addresses.
We consider the major classes of file systems and explore the
challenges each file system type encounters in achieving these two goals.

\newcommand{\addalphaplot}[3]
{
  \addplot[color=#2, line width=0.75pt, mark=#3]
  table[x=read_size_bytes, y expr=\thisrow{read_size_bytes} * \thisrow{num_reads} / \thisrow{time_seconds} / 1000000] {fast_data/#1_alpha_bw.csv};
}

\begin{figure}
\centering
\begin{tikzpicture}[yscale=0.825, xscale=0.825, trim axis left, trim axis right]
    \begin{axis}[
        width=0.75\columnwidth,
        xlabel style={at={(axis description cs:0.5,-0.05)},anchor=north},
        xlabel={Read size (MiB)}, 
        ylabel={Effective bandwidth (MiB per second)}, 
        xmin=4096,
        xmax=1073741824,
        xtick={4096,16384,65536,262144,1048576,4194304,16777216,67108864,268435456,1073741824},
        xticklabel style={align=center},
        xticklabels={0.004, 0.016, 0.063, 0.25, 1, 4, 16, 64, 256, 1024},
        ytick={0.25,1,4,16,64,256,1024},
        yticklabel style={align=center},
        yticklabels={0.25, 1, 4, 16, 64, 256, 1024},
        xmode=log,
        ymode=log,
        log basis x=2,
        log basis y=2,
        grid=major, 
        scaled x ticks=false,
        scaled y ticks=false,
        legend columns=2,
        legend cell align=left,
        legend pos=south east,
    ]
    \addalphaplot{ssd}{red}{square*}
    \addlegendentry{SSD}
    \addalphaplot{hdd}{blue}{*}
    \addlegendentry{HDD}
    \end{axis}
    \end{tikzpicture}
    \caption{Effective bandwidth vs.\ read size. Higher is better.  
    Even on SSDs, large \ios can yield an order of magnitude more bandwidth 
    than small \ios. Note that both axes use log scale.}
\end{figure}
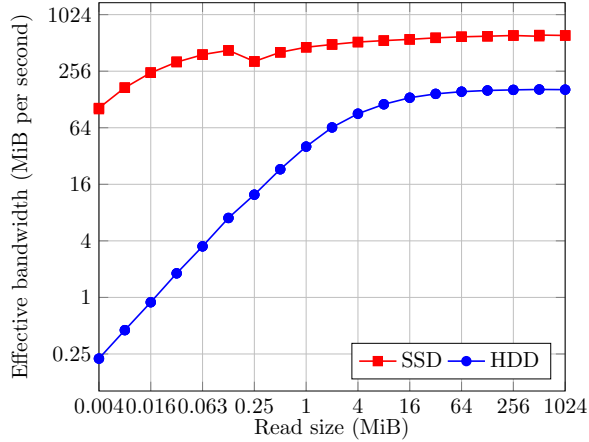~\label{fig:alpha-plot}

\subsecput{model:filesystem}{Allocation Strategies and Intrafile/Interfile Aging}

The major file systems currently in use can be roughly categorized as
B-tree-based, such as \xfs~\cite{SweeneyDoHu96}, \zfs~\cite{BonwickMo08}, 
and \btrfs~\cite{RodehBaMa13};
update-in-place, such as \ext~\cite{CardTsTw94, Tweedie00, MathurCaBh07};
and log-structured, such as \ftwofs~\cite{lee15f2fs}. The research file system
that we consider, \betrfs, is \bet-based.
Each of these fundamental file-system-design categories
have different aging dynamics, discussed in turn below. 
In later sections, we experimentally evaluate file systems from these categories.

\paragraph{B-tree-based file systems.}
The read-aging performance of a B-tree depends on the leaf size. If the leaves are
much smaller than the NTS, then the B-tree will age as the leaves are split and
merged, and thus moved around on the storage device.

Making leaves as large as the  NTS increases \defn{write amplification}, i.e., the
ratio between the amount of data changed and the amount of data written to
storage.  In the extreme case, a single-bit change to a B-tree leaf can cause
the entire leaf to be rewritten.  Thus, B-trees are usually implemented with
small leaves.  Consequently, we expect B-tree-based file systems to age under a wide variety of
workloads.

\secref{git} shows that the read aging of \btrfs is inversely related to
the leaf size, as predicted.  There are, in theory, ways to mitigate
the read aging due to B-tree leaf movements.  For example, the leaves could be
stored in a packed memory array~\cite{BenderDeFa05}.  However, such an
arrangement might well incur an unacceptable performance overhead to keep the
leaves arranged in logical order, and we know of no examples of B-tree
implementation with such leaf-arrangement algorithms.

\paragraph{Update-in-place file systems.}
When data is written once and never moved,
such as in update-in-place file systems like \ext,
sequential order is very difficult to maintain: imagine a workload that writes
two files to disk and then creates files that should logically occur between
them. Without moving one of the original files, sequentiality cannot be maintained.
Such pathological cases abound, and the process is quite
brittle. As noted above, delayed allocation is an attempt to mitigate the
effects of such cases by batching writes and updates before committing them to
the overall structure.

\paragraph{\bet-based file systems.}
\bet-based file systems batch file-system changes in a sequence of
cascading logs, one per node of the tree.  Each time a node overflows, its contents are
flushed to child nodes.  The seeming disadvantage is that data is written
many times, thus increasing the write amplification.  However, each time a node
is modified, it receives many changes, as opposed to B-trees, which might
receive only one change.  Thus, a \bet has asymptotically lower write
amplification than a B-tree.  Consequently, \bets can have much larger nodes, and
typically do in implementation.  \betrfs uses a \bet with 4\mebi\byte{} nodes.

Since 4\mebi\byte{} is around the NTS for our storage devices, we expect \betrfs not to
age---which we verify below.

File systems based on other write-optimized dictionaries,
like log-structured merge trees (LSMs)~\cite{OneilChGa96},
can similarly resist read aging,
depending on the implementation.
As with \bets, it is essential that node sizes match the NTS, the schema reflect
logical access order, and enough writes are batched to avoid heavy write
amplification.

\subsecput{freespacefragfullness}{Free-space Fragmentation and Disk Fullness}
Free-space fragmentation can have a direct effect on write performance,
and an indirect effect on read performance. 
When free space is fragmented, the filesystem must
choose between scattering new data among the existing free-space fragments 
or migrating old data to coalesce free-space fragments.
Both choices come with a cost.
If a filesystem fragments
incoming writes, then the free-space fragmentation gets
turned into regular intra- and inter-file fragmentation, \textit{i.e.,} read aging.
A fragmented write is also slower than when free space is unfragmented,
as one write is split into discrete \ios.
If the file system compacts the free space by moving data, 
the compaction introduces write amplification that slows the write operation.
In either case, free-space fragmentation degrades write performance.

Note that intra- and inter-file fragmentation can exacerbate free-space
fragmentation, and vice versa: fragmented files, when deleted, produce
fragmented free space.

As devices become fuller, managing free-space fragmentation becomes more difficult.
If the file system coalesces free-space fragments,
the cost of coalescing is inversely proportional 
to the fraction of free space available on the disk~\cite{DBLP:journals/talg/BenderFFFG17}.
This is because combining several small free-space fragments into one large fragment
requires moving already-allocated data, which itself needs to be written into free space.
In order to avoid also fragmenting that data,
the allocated data may need to be moved multiple times.
Even on systems that do not coalesce free-space fragments,
fuller disks simply have more allocated objects and less free space.

Free-space fragmentation does not directly impact read performance,
since free space is not actually accessed during a scan.
However, as discussed above,
higher degrees of free-space fragmentation make it harder for file systems to colocate related data;
thus, the relationship between free-space fragmentation and read aging is indirect but very real.

\subsection{Summary}

Because HDDs and most types of SSDs have faster sequential \io than random \io,
file-system fragmentation harms performance, and
the degradation of file-system performance over time due to increased fragmentation is called aging.
There are several different types of fragmentation.
Fragmentation among a single file's allocated blocks
(intra-file fragmentation)
and
fragmentation among the allocated blocks of related files that are in the same directory
(inter-file fragmentation)
directly impact read performance and therefore directly contribute to read aging.
Fragmentation among unallocated blocks,
free-space fragmentation, 
indirectly impacts read performance.
Due to the complex feedback discussed above,
we might expect disk fullness to affect both free-space and intra- and inter-file fragmentation,
and hence affect read and write performance.

To achieve performance that is proportional to the device's available bandwidth,
file systems should perform \ios that are at least as large as their device's natural transfer size.
For commodity HDDs and SSDs, the natural transfer size is large, typically several \mebi\byte.
So that read requests can be satisfied using large sequential \ios,
file systems should colocate related data.
However, preserving data locality requires rewriting data as the system evolves,
which necessarily introduces write amplification.

Thus, for a file-system to avoid aging and maximize long-term performance,
it should dynamically rewrite and group related data,
and when doing so, it should minimize write amplification and avoid fragmenting free space.

Not all file system designs implement this behavior.
The rest of this work examines several such file systems under representative aging workloads 
in order to understand their aging profiles.

\secput{metrics}{Measuring File System Fragmentation}

This section explains the two measures for file system fragmentation
used in our evaluation:
recursive scan (i.e., \texttt{grep -r $\ldots$}) latency
and \defn{dynamic layout score},
a modified form of Smith and Seltzer's \defn{layout score}~\cite{SmithSe97}.
These measures are designed to capture both
intra-file and inter-file fragmentation.

\paragraph{Recursive scan latency.}
The first measure we present is
the wall-clock time required to perform a recursive grep
in the root directory of the file system.
This measure captures the effects of both intra- and inter-file locality,
as a recursive grep scans the contents of both large files and 
large directories containing many related files.
We report search time per unit of data,
normalizing by using \ext's \texttt{du} output.
We will refer to this measure as the \defn{grep test}.

\paragraph{Dynamic layout score.}
Smith and Seltzer's layout score~\cite{SmithSe97}
measures the fraction of blocks in a file
or (in aggregate) a file system 
that is allocated in a contiguous sequence in the logical block space.
We extend this score to capture the dynamic \io patterns of a file system.
During a given workload, 
we observe the \io requests by the file system using \blktrace~\cite{blktrace}, and 
we measure the fraction of the requested blocks that are consecutive.
This approach captures the impact that a file system's placement decisions have on its \io patterns, 
including the impact that placement decisions have on metadata accesses and on accesses that span files.
For a given aging workload,
a high dynamic layout score indicates good data and metadata locality---in other words, an efficient on-disk organization.

One potential shortcoming of this measure is that it does not distinguish
between small and large discontiguities.
Small discontiguities on a hard drive should induce fewer expensive mechanical seeks than large discontiguities in general;
however, factors such as track length, difference in angular placement
and other geometric considerations can complicate this relationship.
A more sophisticated layout measure that penalizes discontiguities
proportional to their magnitude might be more predictive.
We leave this for further research. 
On SSDs, we have found that the length of discontiguities has a smaller effect.  
Thus, we will show that dynamic layout score strongly correlates
with grep test performance on SSDs and
moderately correlates with grep test performance on hard drives.

\paragraph{Measuring fragmentation.}
Though the different forms of fragmentation are interdependent,
we can cleanly measure each fragmentation type at any single moment in time.
We measure free-space fragmentation directly on \ext using \efreefrag~\cite{e2freefrag}.
This tool produces a histogram of the sizes of free extents (unallocated fragments).
Although we do not report the free-space fragmentation on other file systems,
the allocated and free space could be directly inferred by scanning the data with a cold cache
and using a tool such as \blktrace~\cite{blktrace} to observe which blocks are read.
Cold-cache reads can be similarly used to measure intra- and inter-file fragmentation;
the dynamic layout score (described above) captures these fragmentation types.

\paragraph{Write performance and fragmentation.}
When writing a data stream,
a file system's performance is affected by the number of fragments that are written,
since each fragment requires a random \io ---\
writing the same amount of data using fewer fragments will have better performance.
To measure the impact that fragmentation has on writes,
we record the wall-clock latency of new writes.
We find that the aging workloads used in this work are not CPU-bound.

\paragraph{Measuring disk fullness.}
A file system with unlimited free space
is able to apply its ideal allocation strategies without restriction.
As the amount of available free space decreases,
the file system's allocation and placement options become more constrained.
One of the goals of this study is to understand the impact that \defn{disk fullness}
has on file system aging.

Ideally,
we would be able to design experiments that parameterize disk fullness.
One way to do this is to,
for a given disk,
scale a workload to achieve different fullness fractions.
However, scaling the workload necessarily changes the workload, a confounding factor.
Another option is to run the same workload on disks of different sizes. 
However, there are two challenges here:
different physical devices have different hardware specifications,
a confounding factor.
Also, there is a practical limitation on the availability of disk sizes in the market,
restricting measurement granularity.

To standardize our notion of fullness
and to capture the notion of ``restricted placement options'',
we use space pressure as a stand-in for disk fullness.
This choice allows us to
abstract away the differences in individual disk sizes
and run the same workload across different media.

\paragraph{Establishing baselines.}
In order to evaluate the effects of aging and disk fullness,
we need to establish a baseline for comparison that is neither aged nor restricted by space pressure.
Since aging is the result of fragmentation introduced over time by a series of file system operations,
our goal, then, is to create a file system state that has the identical logical contents of 
some aged file system state,
but with an ideal layout, subject to the file system's allocation policies.
Said differently,
if we consider only the \emph{logical} contents of an aged file system at some point in time,
then we want a baseline where those logical contents are organized on disk with 
the maximum locality that the file system's design can achieve.

Note that we never compare the performance of two ``aged'' file system states;
we compare a given ``aged'' file system state against
the optimal layout that the file system could achieve.
This strategy is analogous to competitive analysis~\cite{KarlinMaMc94,DBLP:books/daglib/0023376}
in the theory community.
So, for a target file system state that we wish to evaluate,
we create an unaged baseline as follows.
We allocate a fresh, empty file system on a device that has a single partition spanning its entire logical address space 
(this minimizes space pressure given the physical limitation of finitely-sized devices).
Then, we present this fresh, empty file system 
with the target file system's logical contents in an ordering that corresponds to the files' logical relationships,
as defined by their sort ordering within the namespace hierarchy.
Thus, we are writing the data in an order that corresponds to the
order that data is read during a grep test.

We now argue that a baseline created using this strategy
achieves our goals.
First, by allocating a new file system on an empty partition,
we minimize space pressure.
We cannot truly remove space pressure,
given that no device has unlimited capacity,
but for the experiments and devices used in this study,
this baseline's space pressure is negligible.
Second, our baseline file system state should be ``unaged''.
However, in the process of ``unaging'',
the only things we can control are the operations that we perform and the order that we perform them in.
That is why we ask the fresh, empty file system to write 
files in an order that corresponds to the files' logical sort ordering.
Although the file system's allocation decisions are made based on the file system's current state,
our baseline's write ordering incorporates future knowledge about the final file system state.
Thus, at every point in time,
the file system has the maximum amount of information
to place the files in a way that maximizes the files' locality.

We first run a workload on a small partition
(the ``full disk'' case).
This workload may involve creating, deleting, renaming, writing, etc., files.
It measures the disk fullness and ensures that,
after initial setup, the partition is always above a certain level of fullness.
We record the sequence of operations performed
(such as git pulls or file deletions)
and then replay them on a much larger partition
(the ``empty disk'' case).
Thus the empty and full partitions go through
the exact same sequence of logical filesystem states.

We measure the effect of aging on the full partition, the empty partition, and
a fresh (large) partition to which we have copied the current state (the
``unaged disk'' case). The unaged partition thus provides the baseline
performance of an unaged version of the same filesystem state, and the empty
disk version provides a baseline for the performance of the full disk version.

\secput{setup}{Experimental Setup}

Each experiment compares several file systems: \betrfs, \btrfs, \ext, \ftwofs,
\xfs, and \zfs.  We use the versions of \xfs, \btrfs, \ext and \ftwofs that are
part of the \linuxver kernel, and \zfs 0.6.5.11-1$\_$trusty, downloaded from the
zfsonlinux repository on \url{www.github.com}.  We used BetrFS 0.3 in the 
experiments\footnote{Available at \url{github.com/oscarlab/betrfs}}.
We use default recommended file system settings unless otherwise noted. Lazy
inode table and journal initialization are turned off on \ext, pushing more
work onto file system creation time and reducing experimental noise.

All experimental results are collected on a Intel(R) Xeon(R) CPU with a 4-core 3.00 GHz Intel E3-1220 v6 CPU, 32\gibi\byte{} RAM, a 500\gibi\byte{}, 7200 RPM ATA Toshiba DT01ACA050 disk with a 250\gibi\byte{}, Samsung SSD 860 EVO, with a 512\byte{} block size---both disks used SATA 3.0.
Each file system's block size is set to 4096\byte{}.
Unless otherwise noted, all experiments are cold-cache.

The system runs 64-bit Ubuntu 14.04.6 LTS server with Linux kernel version
\linuxver{} on a bootable USB stick.  All HDD tests besides the mailserver aging benchmark are performed on two 20\gibi\byte{}
partitions located at the outermost region of the drive.
For the SSD tests, we additionally partition the remainder of the drive and
fill it with random data, although we have preliminary data that indicates this
does not affect performance.

\secput{microbenchmarks}{Fragmentation Microbenchmarks}

We present several simple microbenchmarks, each designed around a write/update
pattern for which it is difficult to ensure both fast writes in the moment and
future locality.  These microbenchmarks isolate and highlight the effects of
both intra-file fragmentation and inter-file fragmentation and show the
performance impact aging can have on read performance in the worst cases.

\paragraph{Intrafile Fragmentation.} When a file grows, there may not
be room to store the new blocks with the old blocks on disk, and a
single file's data may become scattered.

Our benchmark creates ten files by first creating each file of an initial size
and then appending between 0 and 100 4\kibi\byte{} chunks of random data in a
round-robin fashion to each of these ten files. In the first round, the initial size of each file is 256\kibi\byte{}, and each entire file is written sequentially, one at a time. 
In subsequent rounds, the number of round-robin chunks increases from 0 to 400\kibi\byte{}, until in the last round, each file is of size 656\kibi\byte{}.
After all the files are written, the caches are flushed by remounting.
This microbenchmark emulates the aging process of multiple files growing in length with time. The file system must allocate space for these files somewhere, but eventually, the files must either be fragmented or moved.

Given that the data set size is small and the test is designed to run in a
short time, an \fsync is performed after each file is written in order to defeat
deferred allocation.

\newcommand{\addrrplot}[2]
{
	\pgfplotstableread{fast_data/intra_#2.csv}\thistable
	\addplot[color=\pgfkeysvalueof{/fs-colors/#1}, style=\pgfkeysvalueof{/agedness-styles/aged}, line width=0.75pt, mark=\pgfkeysvalueof{/fs-marks/#1}, mark repeat=5] table[x=round, y=#1_time] \thistable;
	\addlegendentry{\pgfkeysvalueof{/fs-names/#1}}
}

\newcommand{\addrrlayoutplot}[2]
{
	\pgfplotstableread{fast_data/intra_#2.csv}\thistable
	\addplot[color=\pgfkeysvalueof{/fs-colors/#1}, style=\pgfkeysvalueof{/agedness-styles/aged}, line width=0.75pt, mark=\pgfkeysvalueof{/fs-marks/#1}, mark repeat=5] table[x=round, y=#1_layout] \thistable;
	\addlegendentry{\pgfkeysvalueof{/fs-names/#1}}}

\newcommandx{\startrrplot}[2]
{
  \begin{tikzpicture}[yscale=0.825, xscale=0.825, trim axis left, trim axis right]
    \begin{axis}[
      width=\columnwidth,
	  ylabel style={at={(axis description cs:-0.15,0.5)},anchor=south},
      xlabel={Rounds of 4\kibi\byte{} chunks appended}, 
      ylabel={Grep cost (sec/\gibi\byte{})}, 
      xmin=0,
      xmax=100, 
      ymin=0,  
      ymax=#2,
      grid=major, 
      scaled x ticks=false,
      scaled y ticks=false,
      legend columns=6,
      legend cell align=left,
      legend pos=north west,
      legend to name=rrplotslegend_#1,
      ]
    }

\NewEnviron{rrplot}{\expandafter\startrrplot\BODY
\end{axis}
\end{tikzpicture}
}

\newcommandx{\startrrlayoutplot}[1]
{
  \begin{tikzpicture}[yscale=0.825, xscale=0.825, trim axis left, trim axis right]
    \begin{axis}[
      width=\columnwidth,
	  ylabel style={at={(axis description cs:-0.15,0.5)},anchor=south},
      xlabel={Rounds of 4\kibi\byte{} chunks appended}, 
      ylabel={Dynamic layout score}, 
      xmin=0,
      xmax=100, 
      ymin=0, 
      ymax=1, 
      grid=major, 
      scaled x ticks=false,
      scaled y ticks=false,
      legend columns=6,
      legend cell align=left,
      legend pos=north west,
      legend to name=rrlayoutplotslegend_#1,
      ]
    }

\NewEnviron{rrlayoutplot}{\expandafter\startrrlayoutplot\BODY
\end{axis}
\end{tikzpicture}
}

\newcommand{\rrplotsubcaption}[1]{\label{subfig:rr:#1} Recursive grep cost (\pgfkeysvalueof{/hardware-names/#1}). Lower is better.}
\newcommand{\rrlayoutplotsubcaption}[1]{\label{subfig:rrl:#1} Dynamic layout score (\pgfkeysvalueof{/hardware-names/#1}). Higher is better.}

\begin{figure*}[t]
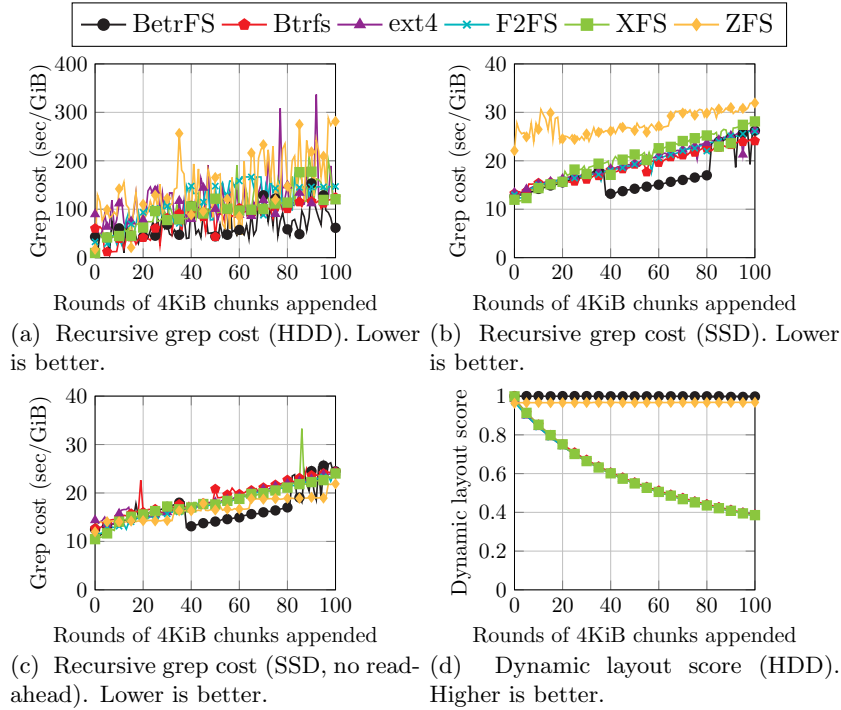

  {\centering
    ~\ref{rrplotslegend_hdd}~\\
	\begin{subfigure}[t]{0.45\columnwidth}
      \centering
      \begin{rrplot}{hdd}{400}
        \addrrplot{betrfs}{hdd}
        \addrrplot{btrfs}{hdd}
        \addrrplot{ext4}{hdd}
        \addrrplot{f2fs}{hdd}
        \addrrplot{xfs}{hdd}
        \addrrplot{zfs}{hdd}
      \end{rrplot}
      \caption{\rrplotsubcaption{hdd}}
    \end{subfigure}
	\begin{subfigure}[t]{0.45\columnwidth}
      \centering
      \begin{rrplot}{ssd}{40}
        \addrrplot{betrfs}{ssd}
        \addrrplot{btrfs}{ssd}
        \addrrplot{ext4}{ssd}
        \addrrplot{f2fs}{ssd}
        \addrrplot{xfs}{ssd}
        \addrrplot{zfs}{ssd}
      \end{rrplot}
      \caption{\rrplotsubcaption{ssd}}
    \end{subfigure}
	\begin{subfigure}[t]{0.45\columnwidth}
      \centering
      \begin{rrplot}{ssd_raoff}{40}
        \addrrplot{betrfs}{ssd_raoff}
        \addrrplot{btrfs}{ssd_raoff}
        \addrrplot{ext4}{ssd_raoff}
        \addrrplot{f2fs}{ssd_raoff}
        \addrrplot{xfs}{ssd_raoff}
        \addrrplot{zfs}{ssd_raoff}
      \end{rrplot}
      \caption{\rrplotsubcaption{ssd_raoff}}
    \end{subfigure}
	 \begin{subfigure}[t]{0.45\columnwidth}
      \centering
       \begin{rrlayoutplot}{hdd}
         \addrrlayoutplot{betrfs}{hdd}
         \addrrlayoutplot{btrfs}{hdd}
		 \addrrlayoutplot{ext4}{hdd}
		 \addrrlayoutplot{f2fs}{hdd}
         \addrrlayoutplot{xfs}{hdd}
         \addrrlayoutplot{zfs}{hdd}
	     \end{rrlayoutplot}
       \caption{\rrlayoutplotsubcaption{hdd}}
     \end{subfigure}
	\caption{\label{fig:micro:intra} Intrafile benchmark: 4\kibi\byte{} chunks are appended round-robin to sequential data to create 10 400\kibi\byte{} files. Dynamic layout scores generally correlate with read performance as measured by the recursive grep test; on an SSD, this effect is hidden by the read-ahead buffer. }
	}
\end{figure*}

The performance of the file systems we tested on an HDD and SSD are summarized in
Figures~\ref{fig:micro:intra}. On HDD, the layout correlates more highly
($-0.85$) with the performance among just \btrfs, \ftwofs, and \xfs, 
as these filesystems' layout scores all degrade over the course of the benchmark.
On SSD, all the file systems excluding \zfs perform similarly 
(note the scale of the y-axis), with \betrfs slightly outperforming 
the rest between rounds 40 and 80 of the benchmark.
In the cases for \btrfs, \ext, \ftwofs, and \xfs, there is a 
strongly negative correlation ($-0.95$) between the grep cost 
and the dynamic layout score.  For \betrfs and \zfs, the performance 
is hidden by read-ahead in the OS; \zfs performance is worse than that of the other 
commercial file systems, and \betrfs is consistently outperforming the others, 
as illustrated in Figure~\ref{subfig:rr:ssd}.
Figure~\ref{subfig:rr:ssd_raoff} shows the performance when we disable the read-ahead; 
the performance is highly correlated ($-.93$) with layout score of \btrfs, \ext, \ftwofs, 
and \xfs.  We do note that this relationship on an SSD is still not precise; SSDs 
are sufficiently fast that factors such as CPU time can also have 
a significant effect on performance.

\newcommand{\addsfplot}[2]
{
	\pgfplotstableread{fast_data/inter_#2.csv}\thistable
	\addplot[color=\pgfkeysvalueof{/fs-colors/#1}, style=\pgfkeysvalueof{/agedness-styles/aged}, line width=0.75pt, mark=\pgfkeysvalueof{/fs-marks/#1}, mark repeat=5] table[x=round, y=#1_time] \thistable;
	\addlegendentry{\pgfkeysvalueof{/fs-names/#1}}
}

\newcommand{\addsflayoutplot}[2]
{
  \pgfplotstableread{fast_data/inter_#2.csv}\thistable
  \addplot[color=\pgfkeysvalueof{/fs-colors/#1}, style=\pgfkeysvalueof{/agedness-styles/aged}, line width=0.75pt, mark=\pgfkeysvalueof{/fs-marks/#1}, mark repeat=5]
  table[x=round, y=#1_layout] \thistable;
  \addlegendentry{\pgfkeysvalueof{/fs-names/#1}}
}

\newcommandx{\startsfplot}[2]
{
  \begin{tikzpicture}[yscale=0.825, xscale=0.825, trim axis left, trim axis right]
    \begin{axis}[
      width=\columnwidth,
	  ylabel style={at={(axis description cs:-0.15,0.5)},anchor=south},
	  xlabel={Percentage of files copied out-of-order}, 
      ylabel={Grep cost (sec/GiB)}, 
      xmin=0,
      xmax=100, 
      ymin=0,  
      ymax=#2,
      grid=major, 
      scaled x ticks=false,
      scaled y ticks=false,
      legend columns=3,
      legend cell align=center,
      legend pos=north west,
      legend to name=sfplotslegend_#1,
      ]
    }

\NewEnviron{sfplot}{\expandafter\startsfplot\BODY
\end{axis}
\end{tikzpicture}
}

\newcommandx{\startsflayoutplot}[1]
{
  \begin{tikzpicture}[yscale=0.825, xscale=0.825, trim axis left, trim axis right]
    \begin{axis}[
      width=\columnwidth,
	  ylabel style={at={(axis description cs:-0.15,0.5)},anchor=south},
      xlabel={Percentage of files copied out-of-order}, 
      ylabel={Dynamic layout score}, 
      xmin=0,
      xmax=100, 
      ymin=0, 
      ymax=1, 
      grid=major, 
      scaled x ticks=false,
      scaled y ticks=false,
      legend columns=3,
      legend cell align=center,
      legend pos=north west,
      legend to name=sfplotslegend_#1,
      ]
    }

\NewEnviron{sflayoutplot}{\expandafter\startsflayoutplot\BODY
\end{axis}
\end{tikzpicture}
}

\newcommand{\sfplotsubcaption}[1]{\label{subfig:sf:#1} Recursive grep cost (\pgfkeysvalueof{/hardware-names/#1}). Lower is better.}
\newcommand{\sflayoutplotsubcaption}[1]{\label{subfig:sfl:#1} Dynamic layout score (\pgfkeysvalueof{/hardware-names/#1}). Higher is better.}

\begin{figure}
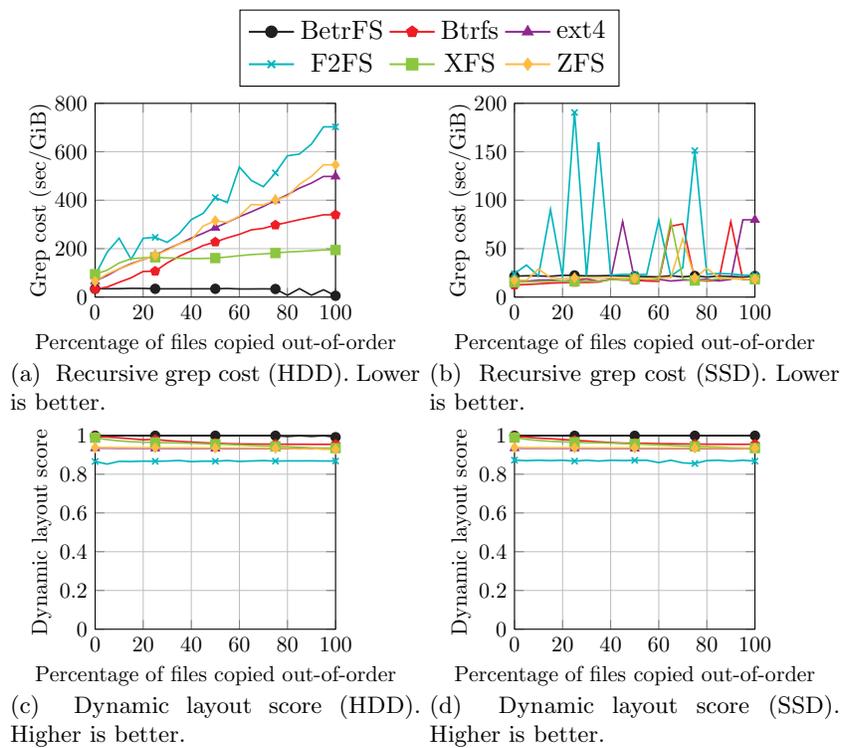

  {\centering
    ~\ref{sfplotslegend_hdd}~\\
	\begin{subfigure}{0.45\columnwidth}\label{subfig:sfhdd}
      \centering
      \begin{sfplot}{hdd}{800}
        \addsfplot{betrfs}{hdd}
        \addsfplot{btrfs}{hdd}
        \addsfplot{ext4}{hdd}
        \addsfplot{f2fs}{hdd}
        \addsfplot{xfs}{hdd}
        \addsfplot{zfs}{hdd}
      \end{sfplot}
      \caption{\sfplotsubcaption{hdd}}
	\end{subfigure}
    \begin{subfigure}{0.45\columnwidth}\label{subfig:sfssd}
      \centering
      \begin{sfplot}{ssd}{200}
        \addsfplot{betrfs}{ssd}
        \addsfplot{btrfs}{ssd}
        \addsfplot{ext4}{ssd}
		\addsfplot{f2fs}{ssd}
        \addsfplot{xfs}{ssd}
        \addsfplot{zfs}{ssd}
      \end{sfplot}
      \caption{\sfplotsubcaption{ssd}}
    \end{subfigure}\\
     \begin{subfigure}{0.45\columnwidth} \label{subfig:sfhddlayout}
       \centering
       \begin{sflayoutplot}{hdd}
         \addsflayoutplot{betrfs}{hdd}
         \addsflayoutplot{btrfs}{hdd}
         \addsflayoutplot{ext4}{hdd}
         \addsflayoutplot{f2fs}{hdd}
         \addsflayoutplot{xfs}{hdd}
         \addsflayoutplot{zfs}{hdd}
       \end{sflayoutplot}
       \caption{\sflayoutplotsubcaption{hdd}}
     \end{subfigure}
     \begin{subfigure}{0.45\columnwidth} \label{subfig:sfssdlayout}
       \centering
       \begin{sflayoutplot}{ssd}
         \addsflayoutplot{betrfs}{ssd}
         \addsflayoutplot{btrfs}{ssd}
         \addsflayoutplot{ext4}{ssd}
         \addsflayoutplot{f2fs}{ssd}
         \addsflayoutplot{xfs}{ssd}
         \addsflayoutplot{zfs}{ssd}
       \end{sflayoutplot}
       \caption{\sflayoutplotsubcaption{ssd}}
     \end{subfigure}
    \caption{\label{fig:micro:inter} Interfile benchmark: All files in the Linux Github repository are replaced by 4\kibi\byte{} random data and copied in varying degrees of order. Dynamic layout scores are predictive of recursive grep performance. }
  }
\end{figure}

\paragraph{Interfile Fragmentation.}\label{sec:interfile} 
Many workloads read
multiple files with some logical relationship, and frequently those files are
placed in the same directory. Interfile fragmentation occurs when files which
are related---in this case being close together in the directory tree---are not
colocated in the LBA space.

We present a microbenchmark to measure the impact of namespace creation order
on interfile locality. It takes a given ``real-life'' file structure, in this
case, the Linux repository obtained from \texttt{github.com}, and copies the
repository's files in a semi-randomized order. This gives us a ``natural'' directory
structure but isolates the effect of file ordering without the influence of
intrafile layout. The benchmark creates a sorted list of the files as well as
a random permutation of a prefix of that list. On each round of the test, the benchmark 
copies a subset of the list, creating directories as needed with 
{\tt cp --parents}. More specifically, on the $n$th round, it swaps the order in which 
a subset of the first $n\%$ of files appearing in the random permutations are copied; all remaining files in the suffix are then copied in order. 
Thus, the first round will be an in-order
copy of the entire list, and subsequent rounds will be copied in progressively more random
order until the last round is a fully random-order copy.

The results of this test are shown in Figure~\ref{fig:micro:inter}. On hard
drive, all the file systems except \betrfs and \xfs show a precipitous
performance decline even if only a small percentage of the files are copied out
of order. \ftwofs often underperforms by at least 100 seconds per \gibi\byte{}, 
compared to any other filesystem, and ends with a grep cost 
20$\times$ that of \betrfs; this is
not entirely unexpected as it is not designed to be used on hard drive. \xfs is
somewhat more stable, although it is 17--36$\times$ slower than drive bandwidth
---as measured with \texttt{hdparm -t}--- throughout the test,  even on an in-order copy.  
\betrfs consistently performs around 1/6 of bandwidth, which by the end of the test is 6$\times$ faster 
than \xfs, and 6--23$\times$ faster than the other file systems. The dynamic layout
scores are moderately correlated with this performance ($-0.68$).

On SSD, all the commercial file systems have sharp increases in grep time 
at several times; this is most pronounced for \ftwofs.
\betrfs is the only file system with stable fast performance; 
it performs at or slightly below the level of all the other file systems, 
but does so consistently, with no spikes in grep time. 

\secput{git}{Application Level Read-Aging: Git}

To measure aging in the ``real-world,'' we create a workload designed to
simulate a developer using git to work on a collaborative project.

Git is a distributed version control system that enables collaborating
developers to synchronize their source code changes.  Git users \defn{pull}
changes from other developers, which then get merged with their own changes.
In a typical workload, a Git user may perform pulls multiple times per day over
several years in a long-running project.  Git can synchronize all types of file
system changes, so performing a Git pull may result in the creation of new
source files, deletion of old files, file renames, and file modifications.  Git
also maintains its own internal data structures, which it updates during pulls.
Thus, Git performs many operations which are similar to those shown in
\secref{microbenchmarks} that cause file system aging.

We present a git benchmark that performs 10,000 pulls from the source Linux git 
repository and places the files in a destination repository, starting from the 
initial commit. Both the source and the destination repository are part of 
the same file system that stays on a single partition. After every 100 pulls, 
the benchmark performs a recursive grep test and computes the file system's 
dynamic layout score. This score is referred to as the dynamic layout score of the aged file system and is compared to the dynamic layout score of an unaged 
file system where the same contents of the aged file system are copied 
to a freshly formatted partition.

\newcommand{\gitgcmodes}{off, on}
\newcommand{\gitDataDir}{fast_data/camera_ready_git}
\newcommand{\gitDataFileName}[4]{\gitDataDir/git_#1_#2_#3_gc_#4results.csv}
\newcommand{\pullCountColumn}{pulls_performed}
\newcommand{\fsSizeColumn}{filesystem_size}
\pgfkeys{
  /agedness-columns/clean/.initial={clean_time},
  /agedness-columns/aged/.initial={aged_time}
}
\pgfkeys{
  /layout-columns/clean/.initial={clean_layout_score},
  /layout-columns/aged/.initial={aged_layout_score}
}

\pgfplotstableread{\gitDataFileName{ext4}{hdd}{4gb}{on}}\extfourgcondata
\pgfplotstableread{\gitDataFileName{ext4}{hdd}{4gb}{off}}\extfourgcoffdata
\pgfplotstableset{create on use/dummy/.style={create col/set list={0,1,2,3,...,10}}}
\pgfplotstablenew[columns={dummy}]{100}\extfourdata
\pgfplotstablecreatecol[copy column from table={\extfourgcondata}{\fsSizeColumn}]{gc_on_fs_size}{\extfourdata}
\pgfplotstablecreatecol[copy column from table={\extfourgcoffdata}{\fsSizeColumn}]{gc_off_fs_size}{\extfourdata}

\newcommand{\addfsplot}[5]{
  \addplot[color=\pgfkeysvalueof{/fs-colors/#1}, style=\pgfkeysvalueof{/agedness-styles/#2}, line width=0.75pt, mark=\pgfkeysvalueof{/fs-marks/#1}, mark repeat=5]
  table[x=#3, y=#4] #5;
}

\newcommand{\fastaddgitplot}[5]
{
  \pgfplotstableread{\gitDataFileName{#1}{#3}{#4}{#5}}\thistable
  \pgfplotstablecreatecol[copy column from table={\extfourdata}{gc_#5_fs_size}]{ext4_fs_size}{\thistable}
  \addplot[color=\pgfkeysvalueof{/fs-colors/#1}, style=\pgfkeysvalueof{/agedness-styles/#2}, line width=0.75pt, mark=\pgfkeysvalueof{/fs-marks/#1}, mark repeat=5] 
  table[x=\pullCountColumn, y expr=\thisrow{\pgfkeysvalueof{/agedness-columns/#2}} / \thisrow{ext4_fs_size} * 1000000] \thistable;
  \addlegendentry{\pgfkeysvalueof{/fs-names/#1} \pgfkeysvalueof{/agedness-names/#2}}
}

\newcommandx{\startgitplot}[3]
{
  \begin{tikzpicture}[yscale=0.825, xscale=0.825, trim axis left, trim axis right]
    \begin{axis}[
      width=\columnwidth,
      xlabel={Pulls accrued}, 
      ylabel={Grep cost (sec/GB)}, 
      xmin=0,
      xmax=10000, 
      ymin=0, restrict y to domain=0:1000,
      grid=major, 
      scaled x ticks=false,
      scaled y ticks=false,
      legend columns=3,
      legend cell align=left,
      legend pos=north west,
      legend to name=gitplotslegend_#1_#2_#3,
      transpose legend,
      ]
    }

\NewEnviron{gitplot}{\expandafter\startgitplot\BODY
\end{axis}
\end{tikzpicture}
}

\newcommand{\gitplotsubcaption}[3]{\label{fig:git:results:#1:#2:#3} \pgfkeysvalueof{/hardware-names/#1}, git garbage collection \pgfkeysvalueof{/git-gc-mode-names/#3}. Lower is better.}

\newcommand{\addlayoutplot}[5]
{
  \pgfplotstableread{\gitDataFileName{#1}{#3}{#4}{#5}}\thistable
  \addplot[color=\pgfkeysvalueof{/fs-colors/#1}, style=\pgfkeysvalueof{/agedness-styles/#2}, line width=0.75pt, mark=\pgfkeysvalueof{/fs-marks/#1}, mark repeat=5] table[x=\pullCountColumn, y expr=\thisrow{\pgfkeysvalueof{/layout-columns/#2}}] \thistable;
  \addlegendentry{\pgfkeysvalueof{/fs-names/#1} \pgfkeysvalueof{/agedness-names/#2}}
}

\newcommandx{\startlayoutplot}[3]
{
  \begin{tikzpicture}[yscale=0.825, xscale=0.825, trim axis left, trim axis right]
    \begin{axis}[
      width=\columnwidth,
      xlabel={Pulls Accrued}, 
      ylabel={Dynamic layout score}, 
      xmin=0,
      xmax=10000, 
      ymin=0, 
      ymax=1, 
      grid=major, 
      scaled x ticks=false,
      scaled y ticks=false,
      legend columns=3,
      legend cell align=left,
      legend pos=north west,
      legend to name=layoutplotslegend_#1_#2_#3,
      transpose legend,
      ]
    }

\NewEnviron{layoutplot}{\expandafter\startlayoutplot\BODY
\end{axis}
\end{tikzpicture}
}

\newcommand{\layoutplotsubcaption}[2]{\label{subfig:git:layout:#1:#2} Dynamic layout score. Git garbage collection \pgfkeysvalueof{/git-gc-mode-names/#2}. Higher is better.}

\newcommand{\buildgitsubplot}[3]{
  \begin{subfigure}{0.48\columnwidth}
    \centering
    \begin{gitplot}{#1}{#2}{#3}
      \foreach \fs in {betrfs, btrfs, ext4, f2fs, xfs, zfs} {
        \foreach \agedness in {clean, aged} {
          \edef\temp{\noexpand\fastaddgitplot{\fs}{\agedness}{#1}{#2}{#3}}
          \temp
        }
      }
    \end{gitplot}
    \caption{\gitplotsubcaption{#1}{#2}{#3}}
  \end{subfigure}
}

\newcommand{\buildlayoutsubplot}[3]{
  \begin{subfigure}{0.48\columnwidth}
    \centering
    \begin{layoutplot}{#1}{#2}{#3}
      \foreach \fs in {betrfs, btrfs, ext4, f2fs, xfs, zfs} {
        \foreach \agedness in {clean, aged} {
          \edef\temp{\noexpand\addlayoutplot{\fs}{\agedness}{#1}{#2}{#3}}
          \temp
        }
      }
    \end{layoutplot}
    \caption{\layoutplotsubcaption{#2}{#3}}
  \end{subfigure}
}

\newcommand{\buildgitplot}[1]{
  \begin{figure*}[p]
    {\centering
      ~\ref{gitplotslegend_hdd_#1_on}~\\
       \buildgitsubplot{hdd}{#1}{on}
	   \hfill
       \buildgitsubplot{hdd}{#1}{off}
	   \\
       \buildgitsubplot{ssd}{#1}{on}
	   \hfill
       \buildgitsubplot{ssd}{#1}{off}
	   \\
       \buildlayoutsubplot{ssd}{#1}{on}
	   \hfill
       \buildlayoutsubplot{ssd}{#1}{off}
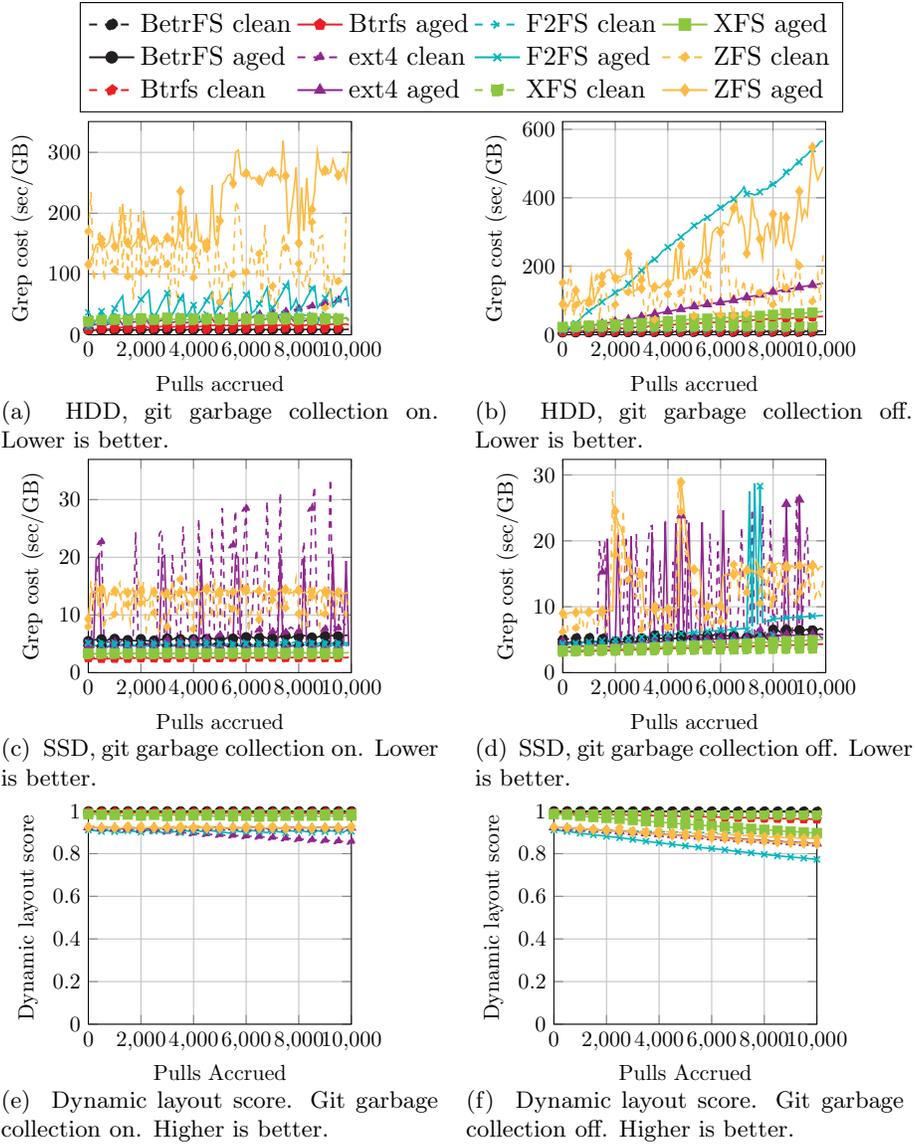
      \caption{\label{fig:git-#1} Git read-aging experimental results. On-disk layout as measured by dynamic layout score is generally predictive of read performance.}
    }
  \end{figure*}
}

\foreach \ps in {20gb} {
  \buildgitplot{\ps}
}

On a hard disk (Figure~\ref{fig:git:results:hdd:20gb:on}), there is a clear
aging trend in all file systems except \betrfs. By the end of the experiment,
all the file systems except \betrfs show performance drops under aging on the
order of at least 2$\times$ relative to their unaged versions. All are 2--31$\times$ worse 
than \betrfs. In all of the experiments in this section, \zfs and \ftwofs age 
considerably more than all other file systems, commensurate with
significantly lower layout scores than the other file systems---indicating less
effective locality in data placement. The overall correlation between grep
performance and dynamic layout score is strongly negative, at $-0.78$. 

On an SSD (Figure~\ref{fig:git:results:ssd:20gb:on}), \btrfs and \xfs show
clear signs of aging, although they converge to a fully aged configuration
after only about 1,000 pulls. While the effect is not as drastic as on HDD, in
all the traditional file systems we see slowdowns of 1.3--2.3$\times$ over \betrfs, which
does not slow down.  In fact, aged \betrfs on the HDD is close to outperform 
all the other aged file systems on an SSD, 
and is close even when they are unaged. Again,
this performance decline is negatively correlated ($-0.59$) with the dynamic
layout scores.

\begin{figure}[t]
  {\centering
    \begin{tikzpicture}[yscale=0.825, xscale=0.825, trim axis left, trim axis right]
      \begin{axis}[
        width=0.6\columnwidth,
		height=0.4\columnwidth,
        xlabel={Average file size}, 
        xmin=4096,
        xmode=log,
	    xmax=65536,
        xtick={4096, 8192, 16384, 32768, 65536},
        xticklabels={4KiB, 8KiB, 16KiB, 32KiB, 64KiB},
        xticklabel style={yshift=-2pt},
        scaled x ticks=false,
        ylabel=Grep cost (sec/GB), 
        ymin=0, 
        ylabel near ticks,
        ymax=400, 
        scaled y ticks=false,
        legend pos=north west,
        ]
        \addplot[color=\pgfkeysvalueof{/fs-colors/ext4}, style=\pgfkeysvalueof{/agedness-styles/clean}, line width=0.0pt, mark=\pgfkeysvalueof{/fs-marks/ext4}] 
        table[x expr=\thisrow{filesystem_size} / \thisrow{file_count} * 1024, y expr=\thisrow{clean_time} / \thisrow{filesystem_size} * 1049600] 
        {fast_data/ext4_git_filesize.csv};
        \addlegendentry{\ext git}

        \addplot[color=LimeGreen, line width=0.75pt, mark=*]
        table[x=file_size, y expr=\thisrow{clean_time} / \thisrow{fs_size} * 1049600]
        {fast_data/ext4_filesize.csv};
        \addlegendentry{\ext interfile}

        \addplot[color=\pgfkeysvalueof{/fs-colors/zfs}, style=\pgfkeysvalueof{/agedness-styles/clean}, line width=0.0pt, mark=\pgfkeysvalueof{/fs-marks/zfs}] 
        table[x expr=\thisrow{filesystem_size} / \thisrow{file_count} * 1024, y expr=\thisrow{clean_time} / \thisrow{filesystem_size} * 1049600] 
        {fast_data/zfs_git_filesize.csv};
        \addlegendentry{\zfs git}

        \addplot[color=Red, line width=0.75pt, mark=x]
        table[x=file_size, y expr=\thisrow{clean_time} / \thisrow{fs_size} * 1049600]
        {fast_data/zfs_filesize.csv};
        \addlegendentry{\zfs interfile}

      \end{axis}
    \end{tikzpicture}
    \caption{\label{fig:git:filesize} 
	Average file size versus unaged grep costs (SSD). Lower is better. 
	Each point on the git lines represents the average file size for the git experiment.
	For each point in the interfile microbenchmark, all files are set to that given size.
	The figure shows a clear relationship between average file size and grep cost.
	\ext performs better on SSD with larger file sizes in both the git and 
	interfile benchmarks.}
}
\end{figure}
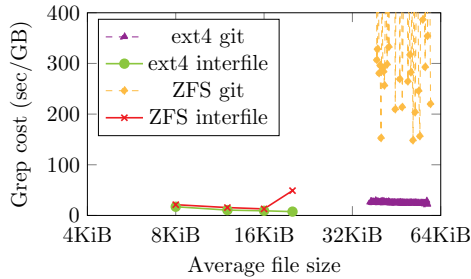

The aged and unaged performance of \ext and \zfs are comparable and slower
than several other file systems.  We believe this is because the average file
size decreases over the course of the test, and these file systems are not as
well-tuned for small files. To test this hypothesis, we constructed
synthetic workloads by copying random data into a randomly constructed repository.
To construct the repository, we started with an empty list of subdirectories, and for 
1000 rounds, we randomly chose a parent directory, into which to insert a child
subdirectory, and added that child to our growing list of parents. Therefore,
in the worst case, our repository would have depth 1000, with a much smaller expected depth.
After creating the empty subdirectories, we randomly determined the locations 
of ~32K files throughout our directory structure.
We then inserted random data of uniform size at these file locations.
This test consists of four rounds: the uniform sizes were 8--20\kibi\byte{}, 
which we increased in increments of 4\kibi\byte{}.
Figure~\ref{fig:git:filesize} shows both the measured average file
size of the git workload (one point is one pull) and the microbenchmark.
Overall, there is a clear relationship between the average file size and grep cost. 

The zig-zag pattern in the graphs is created by an automatic garbage collection
process in Git. Once a certain number of ``loose objects'' are created (in git
terminology), many of them are collected and compressed into a ``pack.'' At the
file system level, this corresponds to merging numerous small files into a
single large file.  According to the Git manual, this process is designed to
``reduce disk space and increase performance'', so this is an example of an
application-level attempt to mitigate file system aging. If we turn off the git
garbage collection, as shown in Figures~\ref{fig:git:results:hdd:20gb:off},
\ref{fig:git:results:ssd:20gb:off} and \ref{subfig:git:layout:20gb:off}, the
effect of aging is even more pronounced, and the zig-zags essentially disappear.

On both the HDD and SSD, the same patterns emerge as with garbage collection
on, but exacerbated: \ftwofs aging is by far the most extreme.  \zfs ages
considerably high on the HDD, but not on the SSD.  \zfs and \ext perform
worse than the other file systems (except \ftwofs aged) on SSD, 
but do not age following a particular pattern. 
\xfs and \btrfs both aged significantly, around 2$\times$ each, and
\betrfs has strong, level performance in both aged and clean states. This performance
correlates with dynamic layout score both on SSD ($-0.78$) and moderately so on
HDD ($-0.54$).

We note that this analysis, both of the microbenchmarks and of the git
workload, runs counter to the commonly held belief that locality is solely an issue on the
hard drive. While the random read performance of solid state drives does
somewhat mitigate the aging effects, aging clearly has a major performance impact.

\paragraph{Git Workload with Warm Cache.}
The tests we have presented so far have all been performed with a cold cache,
so that they more or less directly test the performance of the file systems'
on-disk layout under various aging conditions. In practice, however, some data
will be in cache, and so it is natural to ask how much the layout choices that
the file system makes will affect the overall performance with a warm cache.

We evaluate the sensitivity of the git workloads to varying amounts of system
RAM.  We use the same procedure as above, except that we do not flush any
caches or remount the hard drive between iterations.  This test is performed on
a hard drive with git garbage collection off.  The size of the data on disk is
initially about 4.47\gibi\byte{} and grows throughout the test to approximately 5.2\gibi\byte{}.

\newcommand{\addwcplot}[2]
{
  \pgfplotstableread{fast_data/#2_warm_cache.csv}\thistable
  \addplot[color=\pgfkeysvalueof{/rs-colors/#1}, line width=0.75pt, mark=\pgfkeysvalueof{/rs-marks/#1}, mark repeat=5] table[x=pulls, y=#1_time] \thistable;
  \addlegendentry{\pgfkeysvalueof{/ram-sizes/#1}}
}

\begin{figure}
  {\centering
    ~\ref{wcplotslegend}~\\
	\begin{subfigure}{0.48\columnwidth}
	\centering
	\hspace*{1em}
	 \begin{tikzpicture}[yscale=0.825, xscale=0.825, trim axis left, trim axis right]
	  \begin{axis}[
      width=\columnwidth,
      xlabel={Pulls Accrued}, 
	  ylabel style={align=center},
		  ylabel={(a) \ext\quad Grep cost (sec/GB)}, 
      xmin=0,
      xmax=10000, 
      ymin=0, 
      grid=major, 
      scaled x ticks=false,
      scaled y ticks=false,
      legend columns=4,
      legend cell align=left,
      legend pos=north west,
      legend to name=wcplotslegend,
      transpose legend,
      ]
	  \addwcplot{768mb}{ext4}
	  \addwcplot{1024mb}{ext4}
	  \addwcplot{1280mb}{ext4}
	  \addwcplot{1536mb}{ext4}
	  \addwcplot{2048mb}{ext4}
	  \addwcplot{cold}{ext4}
	  \addwcplot{cold_clean}{ext4}
	\end{axis}
	\end{tikzpicture}
	\end{subfigure}
	\begin{subfigure}{0.48\columnwidth}
	\centering
	\hspace*{1em}
	 \begin{tikzpicture}[yscale=0.825, xscale=0.825, trim axis left, trim axis right]
	  \begin{axis}[
     width=\columnwidth,
     xlabel={Pulls Accrued}, 
	  ylabel style={align=center},
	  ylabel={\vspace*{4em}(b) \btrfs\quad Grep cost (sec/GB)}, 
     xmin=0,
     xmax=10000, 
     ymin=0, 
     grid=major, 
     scaled x ticks=false,
     scaled y ticks=false,
     legend columns=4,
     legend cell align=left,
     legend pos=north west,
     legend to name=wcplotslegend,
     transpose legend,
     ]
	  \addwcplot{768mb}{btrfs}
	  \addwcplot{1024mb}{btrfs}
	  \addwcplot{1280mb}{btrfs}
	  \addwcplot{1536mb}{btrfs}
	  \addwcplot{2048mb}{btrfs}
	  \addwcplot{cold}{btrfs}
	  \addwcplot{cold_clean}{btrfs}
	\end{axis}
	\end{tikzpicture}
	\end{subfigure}
  \begin{subfigure}{0.48\columnwidth}
  \centering
  \hspace*{1em}
   \begin{tikzpicture}[yscale=0.825, xscale=0.825, trim axis left, trim axis right]
    \begin{axis}[
     width=\columnwidth,
     xlabel={Pulls Accrued}, 
    ylabel style={align=center},
    ylabel={\vspace*{4em}(b) \xfs\quad Grep cost (sec/GB)}, 
     xmin=0,
     xmax=10000, 
     ymin=0, 
     grid=major, 
     scaled x ticks=false,
     scaled y ticks=false,
     legend columns=4,
     legend cell align=left,
     legend pos=north west,
     legend to name=wcplotslegend,
     transpose legend,
     ]
    \addwcplot{768mb}{xfs}
    \addwcplot{1024mb}{xfs}
    \addwcplot{1280mb}{xfs}
    \addwcplot{1536mb}{xfs}
    \addwcplot{2048mb}{xfs}
    \addwcplot{cold}{xfs}
    \addwcplot{cold_clean}{xfs}
  \end{axis}
  \end{tikzpicture}
  \end{subfigure}
	\begin{subfigure}{0.48\columnwidth}
	\centering
	\hspace*{1em}
	 \begin{tikzpicture}[yscale=0.825, xscale=0.825, trim axis left, trim axis right]
	  \begin{axis}[
      width=\columnwidth,
      xlabel={Pulls Accrued}, 
	  ylabel style={align=center},
	  ylabel={\vspace*{4em}(b) \zfs\quad Grep cost (sec/GB)}, 
      xmin=0,
      xmax=10000, 
      ymin=0, 
      grid=major, 
      scaled x ticks=false,
      scaled y ticks=true, 
      legend columns=4,
      legend cell align=left,
      legend pos=north west,
      legend to name=wcplotslegend,
      transpose legend,
      ]
	  \addwcplot{768mb}{zfs}
	  \addwcplot{1024mb}{zfs}
	  \addwcplot{1280mb}{zfs}
	  \addwcplot{1536mb}{zfs}
	  \addwcplot{2048mb}{zfs}
	  \addwcplot{cold}{zfs}
	  \addwcplot{cold_clean}{zfs}
	\end{axis}
	\end{tikzpicture}
	\end{subfigure}
	\caption{\label{fig:git:warmcache} 
	Grep costs as a function of system RAM and the number of git pulls
        for
        \ext(top left), \btrfs(top right), 
	\xfs(bottom left), \zfs(bottom right).
        Lower is better.
	Note that the file systems' warm cache performances
	are generally worse than their unaged cold cache performances.}
}
\end{figure}
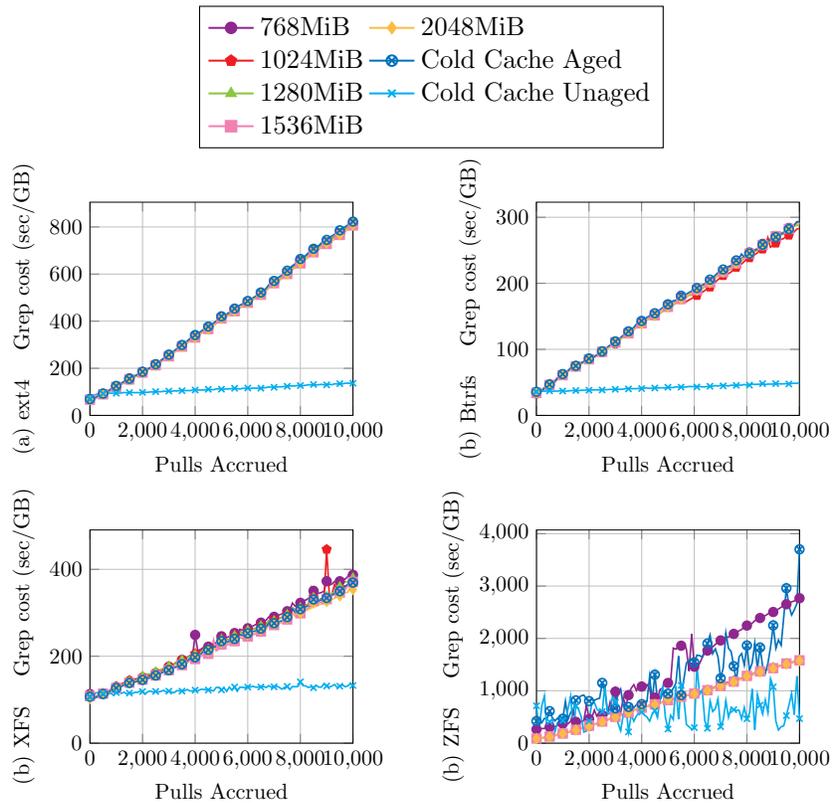

The results are summarized in \figref{git:warmcache}. 
We present data for \ext, \btrfs, \xfs, and \zfs.
\betrfs is a research prototype and unstable under memory pressure; although we 
plan to fix these issues in the future, we omit this comparison.
In general, when the caches are warm and there is sufficient memory to keep all
the data in cache, then the read is very fast. However, as soon as there is no
longer sufficient memory, the performance of the aged file system with a warm
cache is generally worse than unaged with a cold cache.  In general, unless all
data fits into DRAM, a good layout matters more than a having a warm cache.
 
\paragraph{Btrfs Node-Size Trade-Off.}
\Btrfs allows users to specify the node size of its metadata B-tree 
at creation time. Because small files are stored in the metadata
B-tree, a larger node size results in a less fragmented file system, at a cost of
more expensive metadata updates.

We present the git test with a 4\kibi\byte{} node size, the default setting, as well as 8\kibi\byte{}, 16\kibi\byte{}, 32\kibi\byte{}, and 64\kibi\byte{} (the maximum).  
\figref{btrfsGrepTime:results:hdd:20gb:off} shows similar
performance graphs to \figref{git-20gb}, one line for each node size.  The 4\kibi\byte{}
node size has the worst read performance by the end of the test, and the
performance consistently improves as we increase the node size all the way to
64\kibi\byte{}.  \figref{btrfsBlocksWritten:results:hdd:20gb:off} plots the number
of 4\kibi\byte{} blocks written to disk between each test (within the 100 pulls).  As
expected, the 64\kibi\byte{} node size writes the maximum number of blocks and the 4\kibi\byte{}
node writes the least.  We thus demonstrate---as predicted by our model---that
aging is reduced by a larger block size, but at the cost of write amplification.

\pgfkeys{
  /btrfs-nodesize-colors/4k/.initial=black,
  /btrfs-nodesize-colors/8k/.initial=red,
  /btrfs-nodesize-colors/16k/.initial=blue,
  /btrfs-nodesize-colors/32k/.initial=LimeGreen,
  /btrfs-nodesize-colors/64k/.initial=Dandelion
}

\pgfkeys{
  /btrfs-nodesize-marks/4k/.initial=*,
  /btrfs-nodesize-marks/8k/.initial=pentagon*,
  /btrfs-nodesize-marks/16k/.initial=triangle*,
  /btrfs-nodesize-marks/32k/.initial=square*,
  /btrfs-nodesize-marks/64k/.initial=diamond*
}

\pgfkeys{
  /btrfs-nodesize-names/4k/.initial=4Kib,
  /btrfs-nodesize-names/8k/.initial=8Kib,
  /btrfs-nodesize-names/16k/.initial=16Kib,
  /btrfs-nodesize-names/32k/.initial=32Kib,
  /btrfs-nodesize-names/64k/.initial=64Kib
}

\newcommand{\btrfsNSDataDir}{fast_data/btrfs_nodesize}
\newcommand{\btrfsNSDataFileName}[4]{\btrfsNSDataDir/btrfs_#4_#2_#1_gc_#3results.csv}
\newcommand{\btrfsNSFSSizeColumn}{fs_size}
\newcommand{\btrfsNSBlocksWrittenColumn}{blocks_written}
\newcommand{\btrfsNSGrepTimeColumn}{aged_time}

\newcommand{\addbtrfsBlocksWrittenplot}[4]
{
  \addplot[color=\pgfkeysvalueof{/btrfs-nodesize-colors/#4}, line width=0.75pt, mark=\pgfkeysvalueof{/btrfs-nodesize-marks/#4}, mark repeat=5] 
  table[x expr=100 * \coordindex, y expr=\thisrow{\btrfsNSBlocksWrittenColumn} / 1000]{\btrfsNSDataFileName{#1}{#2}{#3}{#4}};
  \addlegendentry{\pgfkeysvalueof{/btrfs-nodesize-names/#4}}
}

\newcommandx{\startbtrfsblockswrittenplot}[3]
{
  \begin{tikzpicture}[yscale=0.825, xscale=0.825, trim axis left, trim axis right]
    \begin{axis}[
      width=\columnwidth,
      xlabel={Pulls accrued}, 
	ylabel style={align=center}, 
	ylabel={Number of 4KiB blocks\\written (thousands)}, 
      xmin=0,
      xmax=10000, 
      ymin=0, 
      grid=major, 
      scaled x ticks=false,
      scaled y ticks=false,
      legend columns=3,
      legend cell align=left,
      legend pos=north west,
      legend to name=btrfsNSBlocksWrittenplotslegend_#1_#2_#3,
      ]
    }

\NewEnviron{btrfsBlocksWrittenplot}{\expandafter\startbtrfsblockswrittenplot\BODY
\end{axis}
\end{tikzpicture}
}

\newcommand{\btrfsBlocksWrittenplotsubcaption}[3]{\label{fig:btrfsBlocksWritten:results:#1:#2:#3} Results: \pgfkeysvalueof{/hardware-names/#1}, \pgfkeysvalueof{/partition-size-names/#2} partition, git garbage collection \pgfkeysvalueof{/git-gc-mode-names/#3}}

\newcommand{\addbtrfsGrepTimeplot}[4]
{
  \addplot[color=\pgfkeysvalueof{/btrfs-nodesize-colors/#4}, line width=0.75pt, mark=\pgfkeysvalueof{/btrfs-nodesize-marks/#4}, mark repeat=5] 
  table[x expr=100 * \coordindex, y expr=\thisrow{\btrfsNSGrepTimeColumn} * 1000000 / \thisrow{\btrfsNSFSSizeColumn}]{\btrfsNSDataFileName{#1}{#2}{#3}{#4}};
  \addlegendentry{\pgfkeysvalueof{/btrfs-nodesize-names/#4}}
}

\newcommandx{\startbtrfsGrepTimeplot}[3]
{
  \begin{tikzpicture}[yscale=0.825, xscale=0.825, trim axis left, trim axis right]
    \begin{axis}[
      width=\columnwidth,
      xlabel={Pulls Accrued}, 
      ylabel={Grep cost (secs/GB)}, 
      xmin=0,
      xmax=10000, 
      ymin=0, 
      ymax=100, 
      grid=major, 
      scaled x ticks=false,
      scaled y ticks=false,
      legend columns=3,
      legend cell align=left,
      legend pos=north west,
      legend to name=btrfsNSGrepTimeplotslegend_#1_#2_#3,
      ]
    }

\NewEnviron{btrfsGrepTimeplot}{\expandafter\startbtrfsGrepTimeplot\BODY
\end{axis}
\end{tikzpicture}
}

\newcommand{\btrfsGrepTimeplotsubcaption}[3]{\label{fig:btrfsGrepTime:results:#1:#2:#3} Results: \pgfkeysvalueof{/hardware-names/#1}, \pgfkeysvalueof{/partition-size-names/#2} partition, git garbage collection \pgfkeysvalueof{/git-gc-mode-names/#3}}

\pgfkeys{
  /btrfs-nodesize-sizes/4k/.initial=4096,
  /btrfs-nodesize-sizes/8k/.initial=8192,
  /btrfs-nodesize-sizes/16k/.initial=16384,
  /btrfs-nodesize-sizes/32k/.initial=32768,
  /btrfs-nodesize-sizes/64k/.initial=65536
}

\pgfkeys{
  /btrfs-nodesize-trend-colors/blocksWritten/.initial=red,
  /btrfs-nodesize-trend-colors/grepSpeed/.initial=blue
}

\pgfkeys{
  /btrfs-nodesize-trend-marks/blocksWritten/.initial=,
  /btrfs-nodesize-trend-marks/grepSpeed/.initial=*
}

\pgfkeys{
  /btrfs-nodesize-trend-names/blocksWritten/.initial=Blocks Written,
  /btrfs-nodesize-trend-names/grepSpeed/.initial=Grep cost (sec/GB)
}

\pgfkeys{
  /btrfs-nodesize-trend-yaxis-sides/blocksWritten/.initial=left,
  /btrfs-nodesize-trend-yaxis-sides/grepSpeed/.initial=right
}

\pgfplotstableset{
  create on use/grepRate/.style={
    create col/expr={\thisrow{\btrfsNSGrepTimeColumn} * 1000000 / \thisrow{\btrfsNSFSSizeColumn}}
  }
}

\begin{figure}
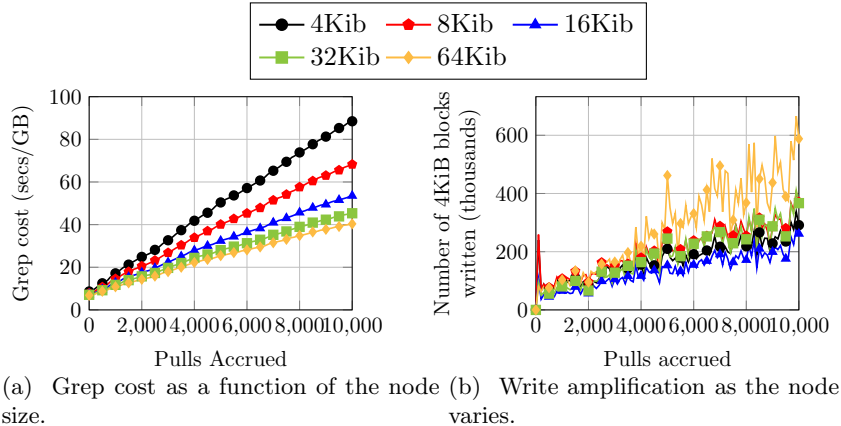

  {\centering
    ~\ref{btrfsNSBlocksWrittenplotslegend_hdd_20gb_off}~\\
    \begin{subfigure}{0.48\columnwidth}
		\centering
      \begin{btrfsGrepTimeplot}{hdd}{20gb}{off}
        \addbtrfsGrepTimeplot{hdd}{20gb}{off}{4k}
        \addbtrfsGrepTimeplot{hdd}{20gb}{off}{8k}
        \addbtrfsGrepTimeplot{hdd}{20gb}{off}{16k}
        \addbtrfsGrepTimeplot{hdd}{20gb}{off}{32k}
        \addbtrfsGrepTimeplot{hdd}{20gb}{off}{64k}
      \end{btrfsGrepTimeplot}
		\caption{\label{fig:btrfsGrepTime:results:hdd:20gb:off}
	Grep cost as a function of the node size.}
    \end{subfigure}
    \begin{subfigure}{0.48\columnwidth}
		\centering
      \begin{btrfsBlocksWrittenplot}{hdd}{20gb}{off}
        \addbtrfsBlocksWrittenplot{hdd}{20gb}{off}{4k}
        \addbtrfsBlocksWrittenplot{hdd}{20gb}{off}{8k}
        \addbtrfsBlocksWrittenplot{hdd}{20gb}{off}{16k}
        \addbtrfsBlocksWrittenplot{hdd}{20gb}{off}{32k}
        \addbtrfsBlocksWrittenplot{hdd}{20gb}{off}{64k}
      \end{btrfsBlocksWrittenplot}
      \caption{\label{fig:btrfsBlocksWritten:results:hdd:20gb:off}
	Write amplification as the node size varies.}
    \end{subfigure}
    \caption{\label{fig:btrfsNodesize} 
	Aging and write amplification on \btrfs, with varying node sizes, 
	under the git aging benchmark. Lower is better.
	Note that a larger node size reduces \btrfs aging
	but increases its write amplification.}
  }
\end{figure}

\secput{mailserver}{Application Level Aging: Mail Server}
In addition to the git workload, we evaluate aging with the Dovecot email server.
Dovecot is configured with the Maildir backend, which stores each message in a file,
and each inbox in a directory.
We simulate 2 users, each having 80 mailboxes
receiving new email, deleting old emails, and searching through their mailboxes. 

A cycle or ``day'' for the mailserver comprises 8,000 operations, where each
operation is equally likely to be an insert or a delete, corresponding to
receiving a new email or deleting an old one. Each email is a string of random
characters, the length of which is uniformly distributed over the range [1,
32K]. Each mailbox is initialized with 1,000 messages, and, because inserts and
deletes are balanced, mailbox size tends to stay around 1,000.  
We simulate the mailserver for 100 cycles and after each cycle we
perform a recursive grep for a random string. Similar to the aforementioned git benchmarks, 
we then copy the partition to a freshly formatted file system, and run a recursive grep.

Figure~\ref{subfig:mailserver_grep_hdd} shows the read costs in seconds 
per \gibi\byte{} of the grep test on hard disk.  
Although the unaged versions of all file systems show consistent
performance over the life of the benchmark,
the aged versions of \ext, \btrfs, \xfs, and \zfs show significant
degradation over time. In particular, aged \ext performance degrades by
4.75$\times$, and is 33$\times$ slower than aged \betrfs.
\xfs slows down by a factor of 10 and \btrfs by a factor of 12.5. \zfs periodically has major dips in read performance, with time for reading 1\gibi\byte{} spiking by up to 100 seconds.
However, the aged version of \betrfs does not slow down. As with the other HDD
experiments, dynamic layout score, as illustrated in 
Figure~\ref{subfig:mailserver_layout_hdd} is moderately correlated 
($-0.64$) with grep cost. 

Figure~\ref{subfig:mailserver_grep_ssd} shows the read costs on solid state drive. 
All unaged file systems show consistent performance, with \ftwofs outperforming 
all others by far. Meanwhile, \betrfs performs comparatively moderately, 
only outperforming \ext and \zfs. Half of the aged file systems are also 
consistent throughout the benchmark, while \betrfs, \btrfs, and \xfs have 
more pronounced degradation over time, with \betrfs degrading by the most and ending with the worst grep performance. More specifically, \betrfs degrades 
by 2.6$\times$, \btrfs degrades by 2.33$\times$, and \xfs degrades by 2.49$\times$. 
Note, however, that no matter the filesystem, the rate at which we read a \gibi\byte{} 
never surpasses 40 seconds, \textit{i.e.}, the range of read times remains small 
across filesystems on SSD. The dynamic layout score, as shown in 
Figure~\ref{subfig:mailserver_layout_ssd}, is more negatively correlated 
($-0.76$) with grep cost on SSD than on HDD. 

\newcommand{\mailserverDataDir}{fast_data/mailserver_aging}
\newcommand{\mailserverDataFileName}[3]{\mailserverDataDir/mailserver_#3_#1_#2.csv}
\newcommand{\mailserverOperationCountColumn}{pulls_performed}
\newcommand{\mailserverFSSizeColumn}{filesystem_size}
\newcommand{\mailserverLayout}[1]{#1_layout_score}

\pgfplotstableread{\mailserverDataFileName{hdd}{20gb}{ext4}}\extfourmsdata

\newcommand{\addmailserverplot}[4]
{
  \pgfplotstableread{\mailserverDataFileName{#3}{#4}{#1}}\thistable
  \pgfplotstablecreatecol[copy column from table={\extfourmsdata}{\mailserverFSSizeColumn}]{ext4_fs_size}{\thistable}
  \addplot[color=\pgfkeysvalueof{/fs-colors/#1}, style=\pgfkeysvalueof{/agedness-styles/#2}, line width=0.75pt, mark=\pgfkeysvalueof{/fs-marks/#1}, mark repeat=5] 
  table[x=\mailserverOperationCountColumn, y expr=\thisrow{\pgfkeysvalueof{/agedness-columns/#2}} / \thisrow{ext4_fs_size} * 1048576] \thistable;
}

\newcommandx{\startmsplot}[2]
{
	\begin{tikzpicture}[xscale = 0.825, yscale = 0.825, trim axis left, trim axis right]
    \begin{axis}[
      width=\columnwidth,
      xlabel={Operations performed}, 
      ylabel={Grep cost (sec/GiB)}, 
      xmin=0,
      xmax=100, 
      ymin=0, 
      grid=major, 
      scaled x ticks=false,
      scaled y ticks=false,
	  legend entries = { \btrfs, \betrfs, \ext, \ftwofs, \xfs, \zfs, aged, unaged},
      legend columns=3,
      legend cell align=center,
      legend pos=north west,
      legend to name=mailserverplotslegend_#1_#2,
      ]
  	\addlegendimage{\pgfkeysvalueof{/fs-colors/btrfs}, mark=\pgfkeysvalueof{/fs-marks/btrfs}}
  	\addlegendimage{\pgfkeysvalueof{/fs-colors/betrfs}, mark=\pgfkeysvalueof{/fs-marks/betrfs}}
  	\addlegendimage{\pgfkeysvalueof{/fs-colors/ext4}, mark=\pgfkeysvalueof{/fs-marks/ext4}}
  	\addlegendimage{\pgfkeysvalueof{/fs-colors/f2fs}, mark=\pgfkeysvalueof{/fs-marks/f2fs}}
  	\addlegendimage{\pgfkeysvalueof{/fs-colors/xfs}, mark=\pgfkeysvalueof{/fs-marks/xfs}}
  	\addlegendimage{\pgfkeysvalueof{/fs-colors/zfs}, mark=\pgfkeysvalueof{/fs-marks/zfs}}
  	\addlegendimage{mark=\pgfkeysvalueof{/fs-marks/ext4}, color=\pgfkeysvalueof{/fs-colors/ext4},  style=\pgfkeysvalueof{/agedness-styles/aged}}
  	\addlegendimage{mark=\pgfkeysvalueof{/fs-marks/ext4}, color=\pgfkeysvalueof{/fs-colors/ext4}, style=\pgfkeysvalueof{/agedness-styles/clean}}
}

\NewEnviron{msplot}{\expandafter\startmsplot\BODY
\end{axis}
\end{tikzpicture}
}

\newcommand{\msplotsubcaption}[2]{\label{fig:ms:results:#1:#2} Results: \pgfkeysvalueof{/hardware-names/#1}, \pgfkeysvalueof{/partition-size-names/#2} partition}

\newcommand{\addmslayoutplot}[4]
{
  \pgfplotstableread{\mailserverDataFileName{#3}{#4}{#1}}\thistable
  \addplot[color=\pgfkeysvalueof{/fs-colors/#1}, style=\pgfkeysvalueof{/agedness-styles/#2}, line width=0.75pt, mark=\pgfkeysvalueof{/fs-marks/#1}, mark repeat=5] 
  table[x=\mailserverOperationCountColumn, y=\mailserverLayout{#2}] \thistable;
}

\newcommandx{\startmslayoutplot}[2]
{
	\begin{tikzpicture}[xscale = 0.825, yscale = 0.825, trim axis left, trim axis right]
    \begin{axis}[
      width=\columnwidth,
      xlabel={Operations performed}, 
      ylabel={Dynamic layout score}, 
      xmin=0,
      xmax=100, 
      ymin=0, 
      grid=major, 
      scaled x ticks=false,
      scaled y ticks=false,
      ]
}

\NewEnviron{mslayoutplot}{\expandafter\startmslayoutplot\BODY
\end{axis}
\end{tikzpicture}
}

\begin{figure}
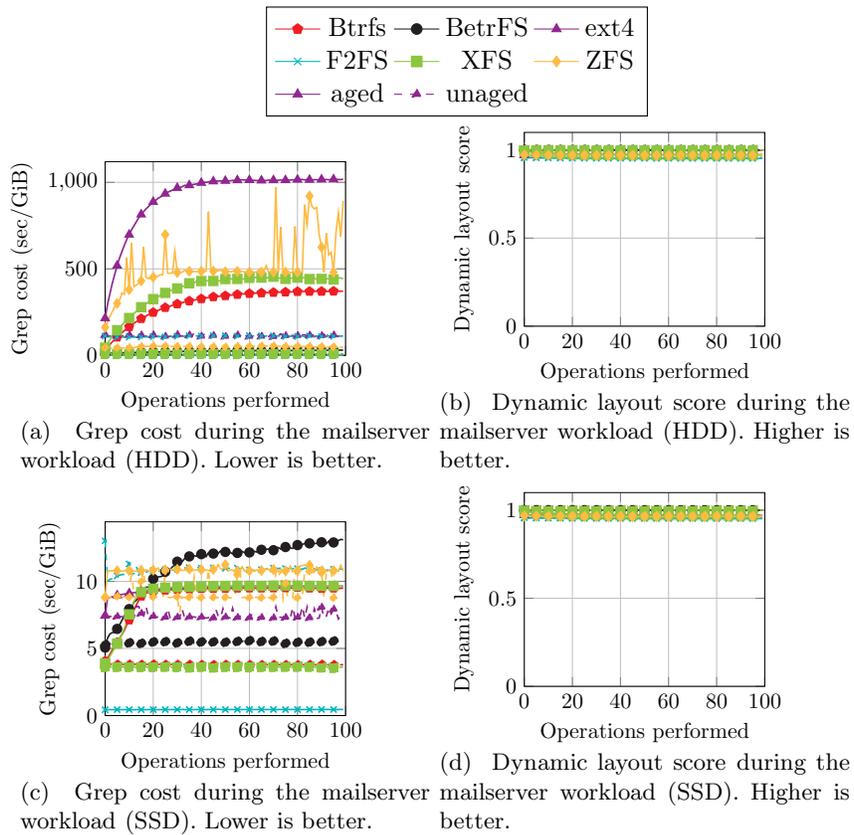

  {\centering
    ~~~~~\ref{mailserverplotslegend_hdd_20gb}\\
    \begin{subfigure}{0.45\columnwidth}
      \centering
      \begin{msplot}{hdd}{20gb}
        \addmailserverplot{betrfs}{clean}{hdd}{20gb}
        \addmailserverplot{betrfs}{aged}{hdd}{20gb}
        \addmailserverplot{btrfs}{clean}{hdd}{20gb}
        \addmailserverplot{btrfs}{aged}{hdd}{20gb}
        \addmailserverplot{ext4}{clean}{hdd}{20gb}
        \addmailserverplot{ext4}{aged}{hdd}{20gb}
        \addmailserverplot{f2fs}{clean}{hdd}{20gb}
        \addmailserverplot{f2fs}{aged}{hdd}{20gb}
        \addmailserverplot{xfs}{clean}{hdd}{20gb}
        \addmailserverplot{xfs}{aged}{hdd}{20gb}
        \addmailserverplot{zfs}{clean}{hdd}{20gb}
        \addmailserverplot{zfs}{aged}{hdd}{20gb}
      \end{msplot}
      \caption{{\label{subfig:mailserver_grep_hdd}} 
      Grep cost during the mailserver workload (HDD). Lower is better.}
    \end{subfigure}
    \begin{subfigure}{0.45\columnwidth}
      \centering
      \begin{mslayoutplot}{hdd}{20gb}
        \addmslayoutplot{betrfs}{clean}{hdd}{20gb}
        \addmslayoutplot{betrfs}{aged}{hdd}{20gb}
        \addmslayoutplot{btrfs}{clean}{hdd}{20gb}
        \addmslayoutplot{btrfs}{aged}{hdd}{20gb}
        \addmslayoutplot{ext4}{clean}{hdd}{20gb}
        \addmslayoutplot{ext4}{aged}{hdd}{20gb}
        \addmslayoutplot{f2fs}{clean}{hdd}{20gb}
        \addmslayoutplot{f2fs}{aged}{hdd}{20gb}
        \addmslayoutplot{xfs}{clean}{hdd}{20gb}
        \addmslayoutplot{xfs}{aged}{hdd}{20gb}
        \addmslayoutplot{zfs}{clean}{hdd}{20gb}
        \addmslayoutplot{zfs}{aged}{hdd}{20gb}
      \end{mslayoutplot}
      \caption{{\label{subfig:mailserver_layout_hdd}} 
      Dynamic layout score during the mailserver workload (HDD). Higher is better.}
    \end{subfigure}
    \begin{subfigure}{0.45\columnwidth}
      \centering
      \begin{msplot}{ssd}{20gb}
        \addmailserverplot{betrfs}{clean}{ssd}{20gb}
        \addmailserverplot{betrfs}{aged}{ssd}{20gb}
        \addmailserverplot{btrfs}{clean}{ssd}{20gb}
        \addmailserverplot{btrfs}{aged}{ssd}{20gb}
        \addmailserverplot{ext4}{clean}{ssd}{20gb}
        \addmailserverplot{ext4}{aged}{ssd}{20gb}
        \addmailserverplot{f2fs}{clean}{ssd}{20gb}
        \addmailserverplot{f2fs}{aged}{ssd}{20gb}
        \addmailserverplot{xfs}{clean}{ssd}{20gb}
        \addmailserverplot{xfs}{aged}{ssd}{20gb}
        \addmailserverplot{zfs}{clean}{ssd}{20gb}
        \addmailserverplot{zfs}{aged}{ssd}{20gb}
      \end{msplot}
      \caption{{\label{subfig:mailserver_grep_ssd}} 
      Grep cost during the mailserver workload (SSD). Lower is better.}
    \end{subfigure}
    \begin{subfigure}{0.45\columnwidth}
      \centering
      \begin{mslayoutplot}{ssd}{20gb}
        \addmslayoutplot{betrfs}{clean}{ssd}{20gb}
        \addmslayoutplot{betrfs}{aged}{ssd}{20gb}
        \addmslayoutplot{btrfs}{clean}{ssd}{20gb}
        \addmslayoutplot{btrfs}{aged}{ssd}{20gb}
        \addmslayoutplot{ext4}{clean}{ssd}{20gb}
        \addmslayoutplot{ext4}{aged}{ssd}{20gb}
        \addmslayoutplot{f2fs}{clean}{ssd}{20gb}
        \addmslayoutplot{f2fs}{aged}{ssd}{20gb}
        \addmslayoutplot{xfs}{clean}{ssd}{20gb}
        \addmslayoutplot{xfs}{aged}{ssd}{20gb}
        \addmslayoutplot{zfs}{clean}{ssd}{20gb}
        \addmslayoutplot{zfs}{aged}{ssd}{20gb}
      \end{mslayoutplot}
      \caption{{\label{subfig:mailserver_layout_ssd}} 
      Dynamic layout score during the mailserver workload (SSD). Higher is better.}
    \end{subfigure}
    \caption{\label{fig:mailserver} Mailserver performance and dynamic layout scores.}
  }
\end{figure}

\section{Full Disk Aging}\label{sec:full-disk-results}

In this section we describe the benchmarks used to generate free-space
fragmentation and the results of running them on several popular filesystems.

\paragraph{Free-space fragmentation microbenchmark (FSFB).}
FSFB is a worst-case microbenchmark, designed to induce severe free-space
fragmentation. FSFB first fills a filesystem with many small files.  Next, it
randomly selects files for deletion and creates a new directory with the same
total size as the deleted files.  Deleting small files creates fragmented free
space, across which the new directory will need to be allocated.

FSFB starts by creating a random directory structure with
1000 directories. Then it creates files by randomly selecting a directory and
creating a file there with size chosen randomly between 1\kibi\byte{} and 150\kibi\byte{}. This process creates the files out-of-directory-order, so that the initial layout is
``pre-aged.'' This process repeats until the file system reaches the target level of fullness.

FSFB then ages the file system through a series of \defn{replacement rounds}.
In a replacement round, 5\% of the files, by size, are removed at random and
then replaced by new files of equivalent total size in a newly created
directory in a random location.

\paragraph{FSFB read aging.}
We run the microbenchmark with a target fullness of 95\% on a 5\gibi\byte{}
partition.  We then age the filesystem for 500 replacement rounds,
performing a grep test every 50 rounds.  We then replay the benchmark
on a 50\gibi\byte{} partition, so that it is at most 10\% full (``empty'').  We
also create an ``unaged'' version by copying the data to a fresh partition.

\newcommand{\addmicroplot}[3]
{
	\addplot[
		color=\pgfkeysvalueof{/fs-colors/#1},
		line width=\plotlinewidth,
		mark=\pgfkeysvalueof{/fs-fullness-marks/#1-#2},
		\pgfkeysvalueof{/fullness-dashes/#2},
		mark options={solid},
	]
	plot[]
	table[
		x=rounds,
		y=grep_cost,
		col sep=tab
	]
	{hs_data/micro_#1_#2_#3.csv};
}

\begin{figure}
	\begin{NoHyper}
	  \centering
	    \ref{legend-micro}
	\vspace*{2pt} \\
	\begin{subfigure}{0.45\columnwidth}
		\begin{tikzpicture}
			\begin{axis}[
				height=0.8\columnwidth,
				width=\columnwidth,
				xlabel=Rounds,
				xlabel near ticks,
				ylabel=Grep Cost (sec/GiB),
				ylabel near ticks,
				xmin=0,
				xmax=500,
				ymin=0,
				ymax=400,
				yticklabel style={rotate=90, xshift={(\tick==400)*-5pt}},
				]
				\addmicroplot{ext4}{95}{hdd}
				\addmicroplot{btrfs}{95}{hdd}
				\addmicroplot{xfs}{95}{hdd}

				\addmicroplot{ext4}{10}{hdd}
				\addmicroplot{btrfs}{10}{hdd}
				\addmicroplot{xfs}{10}{hdd}

				\addmicroplot{ext4}{clean}{hdd}
				\addmicroplot{btrfs}{clean}{hdd}
				\addmicroplot{xfs}{clean}{hdd}
				
			\end{axis}
		\end{tikzpicture}
		\caption{Grep performance under FSFB (HDD).  By the end, all full file
			systems are slower than empty by $1.5-4\times$; \xfs and \btrfs are
			$7\times$ slower empty than unaged.}
		\label{fig:micro-hdd}
	\end{subfigure}
	\begin{subfigure}{0.45\columnwidth}
		\begin{tikzpicture}
			\begin{axis}[
				height=0.8\columnwidth,
				width=\columnwidth,
				xlabel=Rounds,
				xlabel near ticks,
				ylabel=Grep Cost (sec/GiB),
				ylabel near ticks,
				xmin=0,
				xmax=500,
				ymin=0,
				ymax=15,
				yticklabel style={rotate=90, xshift={(\tick==15)*-5pt}},
				legend columns=4,
				transpose legend,
				legend style={nodes={scale=0.9}},
				legend to name=legend-micro
				]
				\addmicroplot{ext4}{95}{ssd}
				\addlegendentry{\ext full}
				\addmicroplot{btrfs}{95}{ssd}
				\addlegendentry{\btrfs full}
				\addmicroplot{xfs}{95}{ssd}
				\addlegendentry{\xfs full}
				\addmicroplot{f2fs}{95}{ssd}
				\addlegendentry{\ftwofs full}

				\addmicroplot{ext4}{10}{ssd}
				\addlegendentry{\ext empty}
				\addmicroplot{btrfs}{10}{ssd}
				\addlegendentry{\btrfs empty}
				\addmicroplot{xfs}{10}{ssd}
				\addlegendentry{\xfs empty}
				\addmicroplot{f2fs}{10}{ssd}
				\addlegendentry{\ftwofs empty}

				\addmicroplot{ext4}{clean}{ssd}
				\addlegendentry{\ext unaged}
				\addmicroplot{btrfs}{clean}{ssd}
				\addlegendentry{\btrfs unaged}
				\addmicroplot{xfs}{clean}{ssd}
				\addlegendentry{\xfs unaged}
				\addmicroplot{f2fs}{clean}{ssd}
				\addlegendentry{\ftwofs unaged}
				
			\end{axis}
		\end{tikzpicture}
		\caption{Grep performance on SSD under FSFB. The full filesystems
			show no discernible slowdown compared to empty, however the empty
			ones are 25-50\% slower than unaged.}
		\label{fig:micro-ssd}
	\end{subfigure}
	\caption{Read performance under FSFB on
		a 95\% full ``full'' disk, a 10\% full ``empty'' disk, and an ``unaged''
		copy. Lower is better.}
	\label{fig:micro}
	\end{NoHyper}
\end{figure}
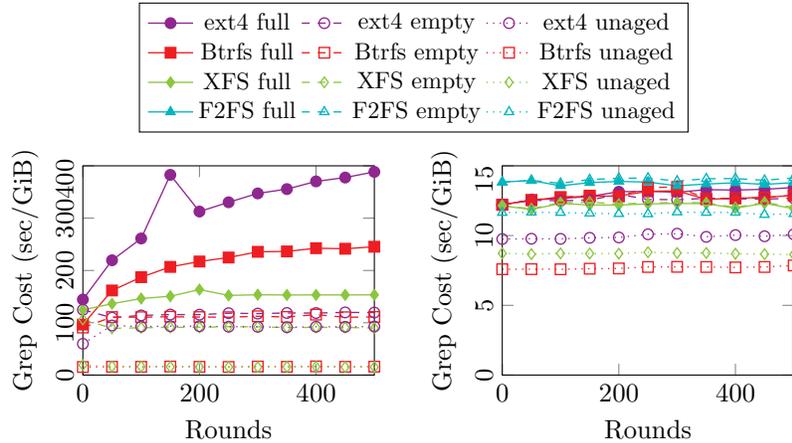

Figure~\ref{fig:micro-hdd} shows the HDD results. All filesystems
are slower in the full-disk case than the empty-disk case. However,
\btrfs and \xfs slow-down far more from unaged to aged than
from empty to full.
\ext, in contrast, only loses read performance under space pressure.

\figref{micro-ssd} shows the SSD results.
The additional read aging from disk fullness is negligible.

\paragraph{FSFB write aging.}
We measure write aging by measuring the wall-clock time to create each
new directory of files during a replacement round.

\newcommand{\addwriteplot}[3]
{
	\addplot[
		color=\pgfkeysvalueof{/fs-colors/#1},
		line width=\plotlinewidth,
		mark=\pgfkeysvalueof{/fs-fullness-marks/#1-#2},
		mark repeat={50},
		\pgfkeysvalueof{/fullness-dashes/#2},
		mark options={solid},
	]
	plot[]
	table[
		x=round,
		y=write_cost,
		col sep=tab
	]
	{hs_data/write_#1_#2_#3.csv};
}

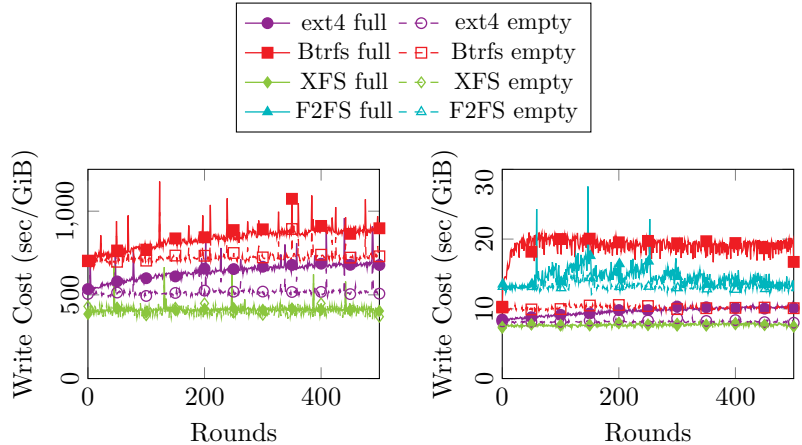
\begin{figure}
	\begin{NoHyper}
	  \centering
	  \ref{legend-write}
	\vspace*{2pt}

	\begin{subfigure}{0.45\columnwidth}
		\begin{tikzpicture}
			\begin{axis}[
				height=0.8\columnwidth,
				width=\columnwidth,
				xlabel=Rounds,
				xlabel near ticks,
				ylabel=Write Cost (sec/GiB),
				ylabel near ticks,
				xmin=0,
				xmax=500,
				ymin=0,
				ymax=1250,
				yticklabel style={rotate=90},
				]
				\addwriteplot{ext4}{full}{hdd}
				\addwriteplot{btrfs}{full}{hdd}
				\addwriteplot{xfs}{full}{hdd}

				\addwriteplot{ext4}{empty}{hdd}
				\addwriteplot{btrfs}{empty}{hdd}
				\addwriteplot{xfs}{empty}{hdd}

			\end{axis}
		\end{tikzpicture}
		\caption{On HDD, \ext slows by 40\% and
		\btrfs by 25\%. \xfs performance is almost unchanged.}
		\label{fig:write-hdd}
	\end{subfigure}
	\begin{subfigure}{0.45\columnwidth}
		\begin{tikzpicture}
			\begin{axis}[
				height=0.8\columnwidth,
				width=\columnwidth,
				xlabel=Rounds,
				xlabel near ticks,
				ylabel=Write Cost (sec/GiB),
				ylabel near ticks,
				xmin=0,
				xmax=500,
				ymin=0,
				ymax=30,
				yticklabel style={rotate=90, xshift={(\tick==30)*-5pt}},
				legend columns=4,
				transpose legend,
				legend style={nodes={scale=0.9}},
				legend to name=legend-write
				]
				\addwriteplot{ext4}{full}{ssd}
				\addlegendentry{\ext full}
				\addwriteplot{btrfs}{full}{ssd}
				\addlegendentry{\btrfs full}
				\addwriteplot{xfs}{full}{ssd}
				\addlegendentry{\xfs full}
				\addwriteplot{f2fs}{full}{ssd}
				\addlegendentry{\ftwofs full}

				\addwriteplot{ext4}{empty}{ssd}
				\addlegendentry{\ext empty}
				\addwriteplot{btrfs}{empty}{ssd}
				\addlegendentry{\btrfs empty}
				\addwriteplot{xfs}{empty}{ssd}
				\addlegendentry{\xfs empty}
				\addwriteplot{f2fs}{empty}{ssd}
				\addlegendentry{\ftwofs empty}

			\end{axis}
		\end{tikzpicture}
		\caption{On SSD, \ext, \xfs, \btrfs, and \ftwofs
		exhibit different behaviors. \ext shows a heavy slowdown.}
			\label{fig:write-ssd}
	\end{subfigure}
	\caption{Write performance under FSFB on a 95\% full ``full''
		disk and a 10\% full ``empty'' disk.  Lower is better.}
	\label{fig:write}
	\end{NoHyper}
\end{figure}

\figref{write-hdd} shows that, on an empty hard drive, none of the
filesystems exhibit any write aging beyond the initial filesystem construction.
When the disk is full, \ext has 40\% higher write costs, \btrfs has 25\% higher
write costs, and \xfs has essentially the same costs. Thus disk fullness does
induce some write aging, but it is an order of magnitude less than read aging
on an empty disk.

On SSDs (\figref{write-ssd}), \xfs is slightly faster when the
disk is full, \ext exhibits a modest 25\% slowdown between the empty
an full cases, \btrfs rapidly loses half its performance in the
full-disk case, and \ftwofs has erratic but generally only slightly slower
performance. Again, except possibly for \btrfs, the performance differences
between an empty and full SSD are smaller than the read aging performance
losses on an empty disk.

As with the read aging effect of disk fullness, space pressure induces a
significant write aging effect, but it is an order of magnitude smaller than
read aging.  The two outlier points were \ext full-disk aging on an HDD and
\btrfs write aging on an SSD.  It might be worth investigating the design
decisions that make these filesystems vulnerable to this workload on a full disk.

\newcommand{\hsaddgitplot}[3]
{
	\addplot[
		color=\pgfkeysvalueof{/fs-colors/#1},
		line width=\plotlinewidth,
		mark=\pgfkeysvalueof{/fs-fullness-marks/#1-#2},
		mark repeat={5},
		\pgfkeysvalueof{/fullness-dashes/#2},
		mark options={solid},
	]
	plot[]
	table[
		x expr={\thisrow{pulls} * 0.001},
		y=grep_cost,
		col sep=tab
	]
	{hs_data/git_#1_#2_#3.csv};
}

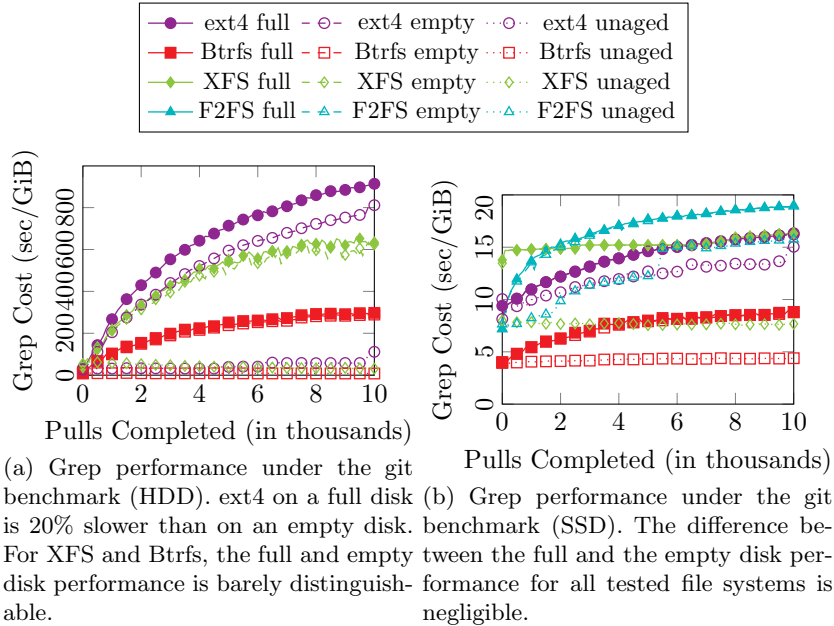
\begin{figure}
	\begin{NoHyper}
	  \centering
	    \ref{legend-git}
	\vspace*{2pt} \\
	\begin{subfigure}{0.45\columnwidth}
		\begin{tikzpicture}
			\begin{axis}[
				height=0.8\columnwidth,
				width=\columnwidth,
				xlabel=Pulls Completed (in thousands),
				xlabel near ticks,
				ylabel=Grep Cost (sec/GiB),
				ylabel near ticks,
				xmin=0,
				xmax=10,
				scaled x ticks=false,
				ymin=0,
				ymax=999,
				yticklabel style={rotate=90},
				]
				\hsaddgitplot{ext4}{full}{hdd}
				\hsaddgitplot{btrfs}{full}{hdd}
				\hsaddgitplot{xfs}{full}{hdd}

				\hsaddgitplot{ext4}{empty}{hdd}
				\hsaddgitplot{btrfs}{empty}{hdd}
				\hsaddgitplot{xfs}{empty}{hdd}
				
				\hsaddgitplot{ext4}{clean}{hdd}
				\hsaddgitplot{btrfs}{clean}{hdd}
				\hsaddgitplot{xfs}{clean}{hdd}
			\end{axis}
		\end{tikzpicture}
		\caption{Grep performance under the git benchmark (HDD). \ext on a full
		disk is 20\% slower than on an empty disk. For \xfs and \btrfs, the full
		and empty disk performance is barely distinguishable.
		}
		\label{fig:git-hdd}
	\end{subfigure}
	\begin{subfigure}{0.45\columnwidth}
		\begin{tikzpicture}
			\begin{axis}[
				height=0.8\columnwidth,
				width=\columnwidth,
				xlabel=Pulls Completed (in thousands),
				xlabel near ticks,
				ylabel=Grep Cost (sec/GiB),
				ylabel near ticks,
				xmin=0,
				xmax=10,
				scaled x ticks=false,
				ymin=0,
				ymax=20,
				yticklabel style={rotate=90, xshift={(\tick==20)*-5pt}},
				legend columns=4,
				transpose legend,
				legend style={nodes={scale=0.9}},
				legend to name=legend-git
				]
				\hsaddgitplot{ext4}{full}{ssd}
				\addlegendentry{\ext full}
				\hsaddgitplot{btrfs}{full}{ssd}
				\addlegendentry{\btrfs full}
				\hsaddgitplot{xfs}{full}{ssd}
				\addlegendentry{\xfs full}
				\hsaddgitplot{f2fs}{full}{ssd}
				\addlegendentry{\ftwofs full}

				\hsaddgitplot{ext4}{empty}{ssd}
				\addlegendentry{\ext empty}
				\hsaddgitplot{btrfs}{empty}{ssd}
				\addlegendentry{\btrfs empty}
				\hsaddgitplot{xfs}{empty}{ssd}
				\addlegendentry{\xfs empty}
				\hsaddgitplot{f2fs}{empty}{ssd}
				\addlegendentry{\ftwofs empty}
				
				\hsaddgitplot{ext4}{clean}{ssd}
				\addlegendentry{\ext unaged}
				\hsaddgitplot{btrfs}{clean}{ssd}
				\addlegendentry{\btrfs unaged}
				\hsaddgitplot{xfs}{clean}{ssd}
				\addlegendentry{\xfs unaged}
				\hsaddgitplot{f2fs}{clean}{ssd}
				\addlegendentry{\ftwofs unaged}
				
			\end{axis}
		\end{tikzpicture}
		\caption{Grep performance under the git benchmark (SSD). 
		The difference between the full and the empty disk performance
		for all tested file systems is negligible.
		}
		\label{fig:git-ssd}
	\end{subfigure}
	\caption{Read performance under the git benchmark. Lower is better.}
	\label{fig:git}
	\end{NoHyper}
\end{figure}

\paragraph{Git benchmark full-disk read aging.}

We also use git as a more representative application benchmark.
We modify the git aging benchmark~\cite{DBLP:conf/fast/ConwayBJJZYBJKP17}, so that it
can be used to keep a disk in a nearly-full steady state.
The git benchmark replays the commit history of the Linux kernel from
\url{github.com}. The benchmark pulls each commit, running a grep test
every 100 commits. 

The challenge to performing the git test on a full disk is that
the repository grows over time.  The disk
starts empty and eventually becomes full, at which time we cannot
pull newer commits.  We overcome this challenge by
maintaining multiple copies of the repository.
We initially fill the disk to 75\% by creating multiple
copies of the initial commit.
Then we update the repositories in a round-robin manner by pulling one more
commit, until a pull fails due to disk fullness.
After the pull fails, at that state of the repository,
the repository is deleted, which frees up space.
Then the process continues.

Every operation is also mirrored on an ``empty'' filesystem and an ``unaged''
version (see \secref{metrics}).  Because this workload is generally CPU-bound
during the pulls, we do not present the effect on write aging.

On an HDD,
there is a big difference between
the empty and unaged versions (\figref{git-hdd}),
commensurate with prior results~\cite{DBLP:conf/fast/ConwayBJJZYBJKP17}.
For \xfs and
\btrfs, the full and empty versions are barely distinguishable. The read cost
for \ext on a full disk is about 20\% greater than on an empty disk.

On SSD, the full and empty lines of all three filesystems are essentially
indistinguishable, shown in \figref{git-ssd}.
On \ext, \ftwofs and, to a lesser extent on \btrfs, the read costs of the unaged
versions drift higher as the benchmark progresses.  This is due to
a smaller average file size.

If free-space aging were a first-order consideration, we would expect it to
consistently create performance degradation in all of these experiments.  In
the git workload, disk fullness has at most a lower-order effect on read
aging than the workload itself.  Its biggest impact was on \ext on
HDD, which added 20\% to the read cost, compared to a 1,200\% increase from the
baseline fragmentation caused by usage with an abundance of space.

\paragraph{Free-Space Fragmentation on \ext}

\begin{figure}
	\begin{NoHyper}
		\begin{subfigure}{\columnwidth}
		  \centering
      \parbox[c]{0.8\columnwidth}{
        \scalebox{0.75}{
			\begin{tikzpicture}
				\begin{axis} [
					height=0.53\columnwidth,
					width=0.55\columnwidth,
					stack plots=y,
					area style,
					enlargelimits=false,
          scaled y ticks={real:1000},
					y tick label style={rotate=90},
          ytick scale label code/.code={},
          ylabel=Space (MiB),
          ylabel near ticks,
		  xlabel=FSFB full,
		  xlabel near ticks,
					xtick=data,
					x tick label style={xshift={(\tick==500)*-8pt}},
				  legend to name=histlegend,
          reverse legend,
					]
					\addplot coordinates
						{ (0, 2254) (50, 4163) (150, 5622) (500, 7201)}
						\closedcycle;
            \addlegendentry{8KiB}
					\addplot coordinates
						{ (0, 9178) (50, 10512) (150, 10678) (500, 6533)}
						\closedcycle;
            \addlegendentry{16KiB}
					\addplot coordinates
						{ (0, 12368) (50, 12687) (150, 16229) (500, 11805)}
						\closedcycle;
            \addlegendentry{32KiB}
					\addplot coordinates
						{ (0, 22170) (50, 19693) (150, 19702) (500, 10394)}
						\closedcycle;
            \addlegendentry{64KiB}
					\addplot coordinates
						{ (0, 30601) (50, 21879) (150, 7644) (500, 2700)}
						\closedcycle;
            \addlegendentry{128KiB}
					\addplot coordinates
						{ (0, 33) (50, 1823) (150, 724) (500, 317)}
						\closedcycle;
            \addlegendentry{256KiB}
					\addplot coordinates
						{ (0, 70) (50, 594) (150, 112) (500, 66)}
						\closedcycle;
            \addlegendentry{512KiB}
					\addplot coordinates
						{ (0, 0) (50, 396) (150, 315) (500, 0)}
						\closedcycle;
            \addlegendentry{1MiB}
					\addplot coordinates
						{ (0, 0) (50, 396) (150, 315) (500, 0)}
						\closedcycle;
            \addlegendentry{2MiB}
					\addplot coordinates
						{ (0, 0) (50, 0) (150, 0) (500, 0)}
						\closedcycle;
            \addlegendentry{4MiB}
					\addplot coordinates
						{ (0, 0) (50, 0) (150, 0) (500, 0)}
						\closedcycle;
            \addlegendentry{8MiB}
					\addplot coordinates
						{ (0, 0) (50, 0) (150, 0) (500, 0)}
						\closedcycle;
            \addlegendentry{16MiB}
					\addplot coordinates
						{ (0, 0) (50, 0) (150, 0) (500, 0)}
						\closedcycle;
            \addlegendentry{32MiB}
					\addplot coordinates
						{ (0, 0) (50, 0) (150, 0) (500, 0)}
						\closedcycle;
            \addlegendentry{64MiB}
					\addplot coordinates
						{ (0, 0) (50, 0) (150, 0) (500, 0)}
						\closedcycle;
            \addlegendentry{128MiB}
					\addplot coordinates
						{ (0, 0) (50, 0) (150, 0) (500, 0)}
						\closedcycle;
            \addlegendentry{256MiB}
					\addplot coordinates
						{ (0, 0) (50, 0) (150, 0) (500, 0)}
						\closedcycle;
            \addlegendentry{512MiB}
					\addplot coordinates
						{ (0, 0) (50, 0) (150, 0) (500, 0)}
						\closedcycle;
            \addlegendentry{1GiB}
					\addplot coordinates
						{ (0, 0) (50, 0) (150, 0) (500, 0)}
						\closedcycle;
            \addlegendentry{2GiB}
				\end{axis}
			\end{tikzpicture}
        }
        \scalebox{0.75}{
			\begin{tikzpicture}
				\begin{axis} [
					height=0.53\columnwidth,
					width=0.55\columnwidth,
					stack plots=y,
					area style,
					enlargelimits=false,
          scaled y ticks={real:1000000},
					y tick label style={rotate=90},
          ytick scale label code/.code={},
          ylabel=Space (GiB),
          ylabel near ticks,
		  xlabel=FSFB empty,
		  xlabel near ticks,
					xtick=data,
					x tick label style={xshift={(\tick==500)*-8pt}}
					]
					\addplot coordinates
						{ (0, 1652) (50, 1148) (150, 1202) (500, 1147)}
						\closedcycle;
					\addplot coordinates
						{ (0, 8547) (50, 5985) (150, 6016) (500, 6037)}
						\closedcycle;
					\addplot coordinates
						{ (0, 36877) (50, 26371) (150, 27251) (500, 24542)}
						\closedcycle;
					\addplot coordinates
						{ (0, 158493) (50, 120245) (150, 121937) (500, 123283)}
						\closedcycle;
					\addplot coordinates
						{ (0, 232787) (50, 191348) (150, 211140) (500, 196905)}
						\closedcycle;
					\addplot coordinates
						{ (0, 262) (50, 307896) (150, 312253) (500, 302493)}
						\closedcycle;
					\addplot coordinates
						{ (0, 0) (50, 554619) (150, 527263) (500, 548016)}
						\closedcycle;
					\addplot coordinates
						{ (0, 0) (50, 734091) (150, 682032) (500, 747240)}
						\closedcycle;
					\addplot coordinates
						{ (0, 801) (50, 828695) (150, 751619) (500, 903743)}
						\closedcycle;
					\addplot coordinates
						{ (0, 0) (50, 847509) (150, 562129) (500, 624818)}
						\closedcycle;
					\addplot coordinates
						{ (0, 0) (50, 568214) (150, 279183) (500, 210517)}
						\closedcycle;
					\addplot coordinates
						{ (0, 0) (50, 145304) (150, 193251) (500, 55163)}
						\closedcycle;
					\addplot coordinates
						{ (0, 0) (50, 0) (150, 255959) (500, 26886)}
						\closedcycle;
					\addplot coordinates
						{ (0, 0) (50, 0) (150, 283104) (500, 0)}
						\closedcycle;
					\addplot coordinates
						{ (0, 194240) (50, 317932) (150, 503809) (500, 284837)}
						\closedcycle;
					\addplot coordinates
						{ (0, 0) (50, 0) (150, 41327) (500, 34884)}
						\closedcycle;
					\addplot coordinates
						{ (0, 259061) (50, 90080) (150, 90080) (500, 90080)}
						\closedcycle;
					\addplot coordinates
						{ (0, 221152) (50, 221152) (150, 473162) (500, 1133286)}
						\closedcycle;
					\addplot coordinates
						{ (0, 10516958) (50, 6669456) (150, 6305456) (500, 6305456)}
						\closedcycle;
				\end{axis}
			\end{tikzpicture}
      }\\
      \scalebox{0.75}{
			\begin{tikzpicture}
				\begin{axis} [
					height=0.53\columnwidth,
					width=0.55\columnwidth,
					stack plots=y,
					area style,
					enlargelimits=false,
          scaled y ticks={real:1000},
					y tick label style={rotate=90},
          ytick scale label code/.code={},
          ylabel=Space (MiB),
          ylabel near ticks,
		  xlabel=git full,
		  xlabel near ticks,
					xtick=data,
					x tick label style={xshift={(\tick==10)*-6pt}}
					]
					\addplot coordinates
						{ (0, 0) (1, 771) (3.5, 5032) (10, 6197)}
						\closedcycle;
					\addplot coordinates
						{ (0, 0) (1, 1627) (3.5, 12705) (10, 19238)}
						\closedcycle;
					\addplot coordinates
						{ (0, 0) (1, 3782) (3.5, 26332) (10, 42216)}
						\closedcycle;
					\addplot coordinates
						{ (0, 0) (1, 4969) (3.5, 46771) (10, 76504)}
						\closedcycle;
					\addplot coordinates
						{ (0, 0) (1, 5932) (3.5, 53483) (10, 75101)}
						\closedcycle;
					\addplot coordinates
						{ (0, 0) (1, 5898) (3.5, 44619) (10, 64448)}
						\closedcycle;
					\addplot coordinates
						{ (0, 0) (1, 8089) (3.5, 42021) (10, 62802)}
						\closedcycle;
					\addplot coordinates
						{ (0, 0) (1, 7854) (3.5, 43227) (10, 65624)}
						\closedcycle;
					\addplot coordinates
						{ (0, 0) (1, 14870) (3.5, 31448) (10, 68151)}
						\closedcycle;
					\addplot coordinates
						{ (0, 0) (1, 6270) (3.5, 25415) (10, 63685)}
						\closedcycle;
					\addplot coordinates
						{ (0, 0) (1, 6888) (3.5, 5627) (10, 45970)}
						\closedcycle;
					\addplot coordinates
						{ (0, 0) (1, 0) (3.5, 0) (10, 19767)}
						\closedcycle;
					\addplot coordinates
						{ (0, 0) (1, 0) (3.5, 0) (10, 12406)}
						\closedcycle;
					\addplot coordinates
						{ (0, 31031) (1, 0) (3.5, 0) (10, 0)}
						\closedcycle;
					\addplot coordinates
						{ (0, 53983) (1, 16797) (3.5, 16512) (10, 16388)}
						\closedcycle;
					\addplot coordinates
						{ (0, 36933) (1, 0) (3.5, 0) (10, 0)}
						\closedcycle;
					\addplot coordinates
						{ (0, 163327) (1, 126626) (3.5, 119730) (10, 114681)}
						\closedcycle;
					\addplot coordinates
						{ (0, 0) (1, 0) (3.5, 0) (10, 0)}
						\closedcycle;
					\addplot coordinates
						{ (0, 0) (1, 0) (3.5, 0) (10, 0)}
						\closedcycle;
				\end{axis}
			\end{tikzpicture}
      }
      \scalebox{0.75}{
			\begin{tikzpicture}
				\begin{axis} [
					height=0.53\columnwidth,
					width=0.55\columnwidth,
					stack plots=y,
					area style,
					enlargelimits=false,
          scaled y ticks={real:1000000},
					y tick label style={rotate=90},
          ytick scale label code/.code={},
          ylabel=Space (GiB),
          ylabel near ticks,
		  xlabel=git empty,
		  xlabel near ticks,
					xtick=data,
					x tick label style={xshift={(\tick==10)*-6pt}}
					]
					\addplot coordinates
						{ (0, 4) (1, 353) (3.5, 1049) (10, 95)}
						\closedcycle;
					\addplot coordinates
						{ (0, 10) (1, 1066) (3.5, 1800) (10, 109)}
						\closedcycle;
					\addplot coordinates
						{ (0, 5) (1, 1624) (3.5, 2868) (10, 128)}
						\closedcycle;
					\addplot coordinates
						{ (0, 8) (1, 2372) (3.5, 4852) (10, 262)}
						\closedcycle;
					\addplot coordinates
						{ (0, 0) (1, 3018) (3.5, 7796) (10, 490)}
						\closedcycle;
					\addplot coordinates
						{ (0, 44) (1, 2362) (3.5, 10805) (10, 669)}
						\closedcycle;
					\addplot coordinates
						{ (0, 86) (1, 366) (3.5, 15376) (10, 852)}
						\closedcycle;
					\addplot coordinates
						{ (0, 0) (1, 1221) (3.5, 20096) (10, 1207)}
						\closedcycle;
					\addplot coordinates
						{ (0, 0) (1, 44751) (3.5, 62927) (10, 15081)}
						\closedcycle;
					\addplot coordinates
						{ (0, 515) (1, 3172) (3.5, 31909) (10, 5785)}
						\closedcycle;
					\addplot coordinates
						{ (0, 2895) (1, 1640) (3.5, 24081) (10, 4304)}
						\closedcycle;
					\addplot coordinates
						{ (0, 0) (1, 11659) (3.5, 25421) (10, 0)}
						\closedcycle;
					\addplot coordinates
						{ (0, 33982) (1, 12462) (3.5, 48625) (10, 9336)}
						\closedcycle;
					\addplot coordinates
						{ (0, 40960) (1, 13387) (3.5, 68401) (10, 49826)}
						\closedcycle;
					\addplot coordinates
						{ (0, 427088) (1, 404279) (3.5, 279530) (10, 138978)}
						\closedcycle;
					\addplot coordinates
						{ (0, 162460) (1, 129012) (3.5, 193518) (10, 250180)}
						\closedcycle;
					\addplot coordinates
						{ (0, 97274) (1, 97274) (3.5, 272318) (10, 544134)}
						\closedcycle;
					\addplot coordinates
						{ (0, 595956) (1, 586558) (3.5, 538468) (10, 1050365)}
						\closedcycle;
					\addplot coordinates
						{ (0, 8155731) (1, 8125541) (3.5, 8094890) (10, 7913682)}
						\closedcycle;
				\end{axis}
			\end{tikzpicture}
      }
      }
      \parbox[c]{0.15\columnwidth}{
        \scalebox{0.65}{\ref{histlegend}}
      }
	\end{subfigure}
	\end{NoHyper}
	\caption{Free space by extent size on \ext for snapshots under FSFB (at 0, 50, 150 and 500 rounds) and git (at 0, 1, 3.5 and 10 thousand pulls). Each bar represents the total free space in extents of the given size.}~\label{fig:freefrag}
\end{figure}
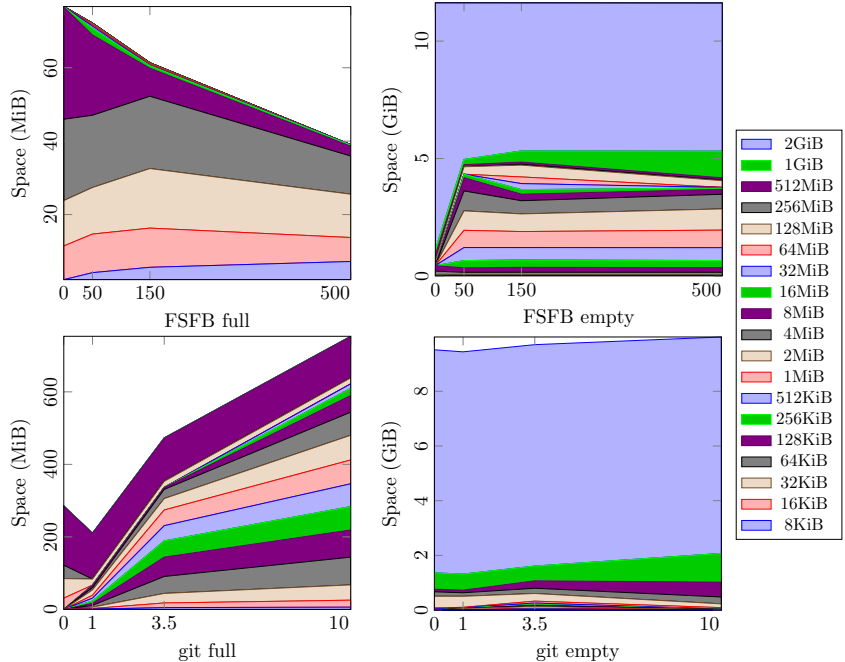

\figref{freefrag} shows the distribution of free-space among
different extent sizes (bucketed into powers of 2),
as reported by e2freefrag~\cite{e2freefrag},
on \ext during our benchmarks.

Both benchmarks create many small free fragments. However, FSFB on
a full disk immediately uses all the large free extents, whereas git
on a full disk and both benchmarks on a empty disk have large free extents available
throughout. Because \ext saw a large performance impact from fullness under
FSFB (\figref{micro}), but not under git (\figref{git}), this suggests that
the availability of large free extents is more important for \ext performance
than the existence of many small free fragments.

\secput{conclusion}{Conclusion}

The experiments above suggest that conventional wisdom on fragmentation, aging,
allocation and file systems is inadequate in several ways.

First, while it may seem intuitive to write data as few times as possible,
writing data only once creates a tension between the logical ordering
of the file system's current state and the potential to make modifications
without disrupting the future order. Rewriting data
multiple times allows the file system to maintain locality.  The overhead
of these multiple writes can be managed by rewriting data in batches, as is done in
write-optimized dictionaries.

For example, in \betrfs, data might be written as many as a
logarithmic number of times, whereas in \ext, it will be written once,
yet \betrfs in general is able to perform as well as or better than an
unaged \ext file system and significantly outperforms aged \ext file
systems.

Second, today's file system heuristics are not able to maintain enough
locality to enable reads to be performed at the disks natural transfer
size.  And since the natural transfer size on a rotating disk is a
function of the seek time and bandwidth, it will tend to increase with
time. Thus we expect this problem to possibly become worse with newer
hardware, not better.

We experimentally confirmed our expectation that non-write-optimized
file systems would age, but we were surprised by how
quickly and dramatically aging impacts performance.
This rapid aging is important: a user's
experience with unaged
file systems is likely so fleeting that they do not notice performance degradation.
Instead, the performance costs of aging
are built into their expectations of file system performance.

Finally, because representative aging is a difficult goal, simulating multi-year workloads,
many research 
papers benchmark on unaged file systems.
Our results indicate that it is relatively easy to quickly drive
a file system into an aged state---even if
this state is not precisely the state of the file system after, say, three
years of typical use---and this degraded state can be easily 
measured.

\secput{related}{Related Work}
Prior work on file system aging can be broadly grouped into three categories:
techniques for artificially inducing aging, for
measuring aging, and for mitigating aging.

\subsection{Creating Aged File Systems}

It takes years to collect years of traces from live systems.
Moreover, traces are large, idiosyncratic, and may contain
sensitive data.  Consequently, researchers have created synthetic benchmarks to
simulate aging.
Once aged, a filesystem can be
profiled using other benchmarking tools to understand how an initial aged state
affects \emph{future} operations.

The seminal work of Smith and Seltzer~\cite{SmithSe97} created a
methodology for 
simulating and measuring aging on a file system---leading
to more representative benchmark results than 
running on a new, empty file system.
The study is based on data  collected
from daily snapshots of more than fifty real file systems from five servers
over durations ranging from one to three years. 
An overarching goal of Smith and Seltzer's work was to evaluate
file systems with representative levels of aging.

Other tools have been subsequently developed for
synthetically aging a file system.
TBBT~\cite{ZhuChCh05} was designed 
to synthetically age a disk
in order to create a starting point for an NFS trace replay.
TBBT first creates a
namespace hierarchy, then interleaves synthetic operations so that
allocations are more fragmented.

The Impressions framework~\cite{AgrawalArAr09} was designed so that
users can synthetically age a 
file system by setting a small number of parameters,
such as the organization of the directory hierarchy.
Impressions also lets users specify a target layout score for the resulting image.

Like Impressions,
Geriatrix is a software tool that generates synthetic aging workloads~\cite{kadekodi18atc}.
Geriatrix is unique in that users can provide aging profiles
to fragment both allocated file blocks and the free space within the file system.
In addition to the Geriatrix tool,
the project contributes a set of built-in aging profiles
and
a repository of aged file system images.

TBBT, Impressions, and Geriatrix all
create file systems with a specific level of fragmentation,
whereas our study identifies realistic
workloads that induce fragmentation. 

\subsection{Quantifying File System Aging}

Smith and Seltzer also introduced a \defn{layout score} for studying aging,
which was used by subsequent studies~\cite{ahn02mascots,AgrawalArAr09}. Their
layout score is the fraction of file blocks that are placed in consecutive
physical locations on the disk. We introduce a variation of this measure, the
\defn{dynamic layout score} in \secref{metrics}.

The \defn{degree of fragmentation} (\defn{DoF}) is used in the study of
fragmentation in mobile devices~\cite{JiChSh16}. DoF is the ratio of the actual
number of extents, or ranges of contiguous physical blocks, to the ideal number
of extents.  Both the layout score and DoF measure how one file is fragmented.

Several studies have reported file system statistics
such as number of files, distributions of
file sizes and types, and organization of file system
namespaces~\cite{AgrawalBoDo07,Downey01,RoselliLoAn01}.
These statistics can inform parameter choices in aging frameworks like TBBT and Impressions~\cite{ZhuChCh05,AgrawalArAr09}.

Ji et al.\ \cite{DBLP:conf/hotstorage/JiCSWLX16}  studied filesystem
fragmentation on mobile devices, confirming that
fragmentation causes performance degradation on mobile devices and that existing
defragmentation techniques are ineffective on mobile devices.

\subsecput{filesystem}{Strategies to Mitigate Aging}

When files are created or extended, blocks must be allocated to store the new
data.  Especially when data is rarely or never relocated, as in an
update-in-place file system like \ext{},
initial block allocation decisions determine
performance over the life of the file system. 

\paragraph{Cylinder or Block Groups.}
FFS~\cite{McKusickJoLe84} introduced the idea of \defn{cylinder groups},
which later evolved into block groups or allocation groups (\xfs).
Each
group maintains information about its inodes and a bitmap of blocks. A
new directory is placed in the cylinder group that contains more than the
average number of free inodes, while inodes and data blocks of files in one
directory are placed in the same cylinder group when possible.

\zfs~\cite{BonwickMo08} is designed to pool
storage across multiple devices~\cite{BonwickMo08}.
\zfs selects from one of a few hundred \defn{metaslabs} on a 
device, based on a weighted calculation of several factors
including minimizing seek distances.
The metaslab with the highest weight is chosen.

In the case of \ftwofs~\cite{lee15f2fs}, a log-structured file system, 
the disk is divided into segments---the granularity at which the log is garbage collected, or cleaned.
The primary locality-related optimization in \ftwofs is that writes
are grouped to improve locality, and dirty segments are filled before
finding another segment to write to. In other words, writes with temporal
locality are more likely to be placed with physical locality.

Groups are a best-effort approach to directory locality:
space is reserved for co-locating files in the same directory,
but when space is exhausted, files in the same directory can be scattered
across the disk.  Similarly, if a file is renamed, it is not physically moved
to a new group.

\paragraph{Extents.} All of the file systems we measure, except \ftwofs and
\betrfs,
allocate space using \defn{extents}, or runs of physically contiguous blocks.
In \ext~\cite{CardTsTw94,Tweedie00, MathurCaBh07}, for example, an extent can be up to 128 MiB.  Extents reduce
bookkeeping overheads (storing a range versus an exhaustive list of blocks).
Heuristics to select larger extents can improve locality of large files.  For
instance, \zfs selects from available extents in a metaslab using a first-fit
policy.

\paragraph{Delayed Allocation.} Most modern file systems, including \ext, \xfs,
\btrfs, and \zfs, implement delayed allocation, where 
logical blocks are not allocated until buffers are written to disk. 
By delaying allocation when a file is growing,
the file system can allocate a larger extent for data appended to the same file.
However, allocations can only be delayed so long without
violating durability and/or consistency requirements; a typical
file system ensures data is dirty no longer than a few seconds.
Thus, delaying an allocation only improves locality inasmuch
as adjacent data is also written on the same time-scale;
delayed allocation alone cannot prevent fragmentation when 
data is added or removed  over larger time-scales.

Application developers may also request a persistent preallocation
of contiguous blocks using \texttt{fallocate}.
To take full advantage of this interface, 
developers must know each file's size in advance.
Furthermore, \texttt{fallocate} can only help intrafile fragmentation; there is currently 
not an
analogous interface to ensure directory locality.

\paragraph{Packing small files and metadata.}
For directories with many small files, an important optimization 
can be to pack the file contents, and potentially metadata,
into a small number of blocks or extents.
\btrfs~\cite{RodehBaMa13} stores metadata of files and directories in
copy-on-write B-trees. Small files are broken into 
one or more fragments, which are packed inside the B-trees.
For small files, the fragments are indexed by object identifier (comparable to inode number);
the locality of a directory with multiple small files
depends upon the proximity of the object identifiers.

\betrfs stores metadata and data as key-value
pairs in two \bets.  Nodes in a \bet are large (2--4 MiB),
amortizing seek costs.  Key/value pairs are packed
within a node by sort-order, and nodes are periodically rewritten,
copy-on-write, as changes are applied in batches.

\betrfs also divides the namespace of the file system into \defn{zones} of a
desired size (512 KiB by default), in order to maintain locality within a
directory as well as implement efficient renames.  Each zone root is either a
single, large file, or a subdirectory of small files.
The key for a file or directory is its relative path to its zone root. The
key/value pairs in a zone are contiguous, thereby maintaining locality.

\paragraph{Defragmentation and Garbage Collection.}
File system defragmentation is a classic aging mitigation technique
traditionally employed on disk-based devices like HDDs,
where LBA fragmentation induces expensive seeks.
Many of the file systems
in this study provide online or offline defragmentation utilities~\cite{Tso15,oracleXFSdefrag, btrfsmanpage,f2fstools},
which can be used to gather each file's blocks and group related data and metadata on disk.
Defragmenters like these that are tightly coupled to specific file system designs
can leverage data structure knowledge and low-level file system APIs to
consolidate logically related data at the (often high) cost of rewriting.

FragPicker~\cite{park21sosp} is a defragmentation tool
that is not tied to any specific file system design or device type;
instead, FragPicker adapts its data rewriting policies based on
a file system's update paradigm, e.g., update-in-place or no-overwrite.
FragPicker's policies attempt to minimize writes---\
which harm newer devices that have limited endurance---\
and focus on reducing request-splitting,
i.e., breaking a request for a logical range of data into multiple block requests to the block \io subsystem.
To accomplish these goals,
FragPicker monitors an application's \io patterns
in an \emph{analysis} phase
and then migrates only data ranges that it predicts will most impact future performance.

Similar to defragmentation,
garbage collection in log-structured file systems~\cite{RosenblumOu92} rewrites and relocates file system data.
The primary goals of garbage collection are to reclaim space and to defragment free space.
However, garbage collection may harm read performance because related
blocks can be moved farther from each other. A recently proposed
defragmentation scheme for log-structure file systems~\cite{ParkKaEo16}
reorders blocks in inode order before writing back to disk.
This can improve locality within a segment, but cannot
address all types of fragmentation, such as scattering a file across segments.

\section*{Acknowledgments}

Part of this work was done while Jannen, Jiao, Porter, Yuan, and Zhan were at
Stony Brook University, and while Bakshi, Conway, and Knorr were at Rutgers University.

We gratefully acknowledge support from a NetApp Faculty Fellowship,
NSF grants
CCF 805476,
CCF 822388,
CCF 1314547,
CNS 1161541,
CNS 1408695,
CNS 1408782, 
CNS 1405641, 
CNS 1409238,
CNS 1755615,
CCF 1439084,
CCF 1725543,
CSR 1763680,
CCF 1716252,
CCF 1617618,
IIS 1247726,
IIS 1247750,
and from VMware.

\bibliographystyle{plain}
\bibliography{all}

\raggedright\newcommand{\noopsort}[1]{} \newcommand{\singleletter}[1]{#1}
  \punt{ Uncomment the following lines for short conference/journal names
  @String{SODA = {SODA}} @String{JACM = {Journal of the ACM}} @String{SPAA =
  {SPAA}} @String{PPoPP = {PPoPP}} @String{PLDI = {PLDI}} @String{STOC =
  {STOC}} @String{FOCS = {FOCS}} @String{ESA = {ESA}} @String{ALP = {Colloquium
  on Automata, Languages, and Programming}} @String{SWAT = {SWAT}}
  @String{JALGO = {Journal of Algorithms}} @String{PODC = {PODC}} @String{LNCS
  = {LNCS}} @String{SUPERCOMP = {Supercomputing}} @String{ICCSE = {Israeli
  Conference on Computer Systems Engineering}} @String{CMD = {Conference on
  Management of Data}} }
\begin{thebibliography}{10}

\bibitem{AgrawalArAr09}
Nitin Agrawal, Andrea~C. Arpaci-Dusseau, and Remzi~H. Arpaci-Dusseau.
\newblock Generating realistic impressions for file-system benchmarking.
\newblock {\em ACM Transactions on Storage}, 5(4), December 2009.

\bibitem{AgrawalBoDo07}
Nitin Agrawal, William~J. Bolosky, John~R. Douceur, and Jacob~R. Lorch.
\newblock A five-year study of file-system metadata.
\newblock {\em Trans. Storage}, 3(3), October 2007.

\bibitem{ahn02mascots}
Woo~Hyun Ahn, Kyungbaek Kim, Yongjin Choi, and Daeyeon Park.
\newblock {DFS:} {A} de-fragmented file system.
\newblock In {\em 10th IEEE International Symposium on Modeling, Analysis and
  Simulation of Computer and Telecommunications Systems (MASCOTS)}, pages
  71--80, 2002.

\bibitem{btrfsmanpage}
Btrfs.
\newblock Wiki section on defragmentation.
\newblock Accessed 10 May 2016.

\bibitem{blktrace}
Jens Axboe, Alan~D.\ Brunelle, and Nathan Scott.
\newblock blktrace(8) - linux man page.
\newblock \url{https://linux.die.net/man/8/blktrace}.

\bibitem{BenderDeFa05}
M.~A. Bender, E.~Demaine, and M.~Farach-Colton.
\newblock Cache-oblivious {B}-trees.
\newblock {\em SIAM Journal on Computing}, 35(2):341--358, 2005.

\bibitem{DBLP:journals/talg/BenderFFFG17}
Michael~A. Bender, Martin Farach{-}Colton, S{\'{a}}ndor~P. Fekete, Jeremy~T.
  Fineman, and Seth Gilbert.
\newblock Cost-oblivious storage reallocation.
\newblock {\em {ACM} Trans. Algorithms}, 13(3):38:1--38:20, 2017.

\bibitem{BonwickMo08}
Jeff Bonwick and B.~Moore.
\newblock {ZFS}: The last word in file systems.
\newblock In {\em SNIA Developers Conference}, Santa Clara, CA, USA, September
  2008.
\newblock Slides at
  \url{https://www.snia.org/sites/default/orig/sdc_archives/2008_presentations/monday/JeffBonwick-BillMoore_ZFS.pdf},
  talk at \url{https://blogs.oracle.com/video/entry/zfs_the_last_word_in}.
  Accessed 10 May 2016.

\bibitem{CardTsTw94}
R\'emy Card, Theodore Ts'o, and Stephen Tweedie.
\newblock Design and implementation of the {S}econd {E}xtended {F}ilesystem.
\newblock In {\em Proceedings of the First {D}utch International Symposium on
  {L}inux}, pages 1--6, Amsterdam, NL, December~8--9 1994.

\bibitem{ChenKoZh09}
Feng Chen, David~A. Koufaty, and Xiaodong Zhang.
\newblock Understanding intrinsic characteristics and system implications of
  flash memory based solid state drives.
\newblock In {\em Proceedings of the Eleventh International Joint Conference on
  Measurement and Modeling of Computer Systems}, SIGMETRICS '09, pages
  181--192, New York, NY, USA, 2009. ACM.

\bibitem{DBLP:conf/fast/ConwayBJJZYBJKP17}
Alexander Conway, Ainesh Bakshi, Yizheng Jiao, William Jannen, Yang Zhan, Jun
  Yuan, Michael~A. Bender, Rob Johnson, Bradley~C. Kuszmaul, Donald~E. Porter,
  and Martin Farach{-}Colton.
\newblock File systems fated for senescence? nonsense, says science!
\newblock In Geoff Kuenning and Carl~A. Waldspurger, editors, {\em USENIX
  FAST}, pages 45--58. {USENIX} Association, 2017.

\bibitem{DBLP:books/daglib/0023376}
Thomas~H. Cormen, Charles~E. Leiserson, Ronald~L. Rivest, and Clifford Stein.
\newblock {\em Introduction to Algorithms, 3rd Edition}.
\newblock {MIT} Press, 2009.

\bibitem{Downey01}
Allen~B. Downey.
\newblock The structural cause of file size distributions.
\newblock In {\em Proceedings of the 2001 ACM SIGMETRICS International
  Conference on Measurement and Modeling of Computer Systems}, SIGMETRICS '01,
  pages 328--329, New York, NY, USA, 2001. ACM.

\bibitem{EsmetBeFa12}
John Esmet, Michael~A. Bender, Martin Farach-Colton, and B.~C. Kuszmaul.
\newblock The {TokuFS} streaming file system.
\newblock In {\em 4th USENIX Workshop on Hot Topics in Storage and File Systems
  (HotStorage 12)}, 2012.

\bibitem{pcworld-ssd-defrag-benchmarks}
Jon Jacobi.
\newblock Fragging wonderful: The truth about defragging your ssd.
\newblock Accessed 25 September 2016.

\bibitem{jannen15betrfs}
William Jannen, Jun Yuan, Yang Zhan, Amogh Akshintala, John Esmet, Yizheng
  Jiao, Ankur Mittal, Prashant Pandey, Phaneendra Reddy, Leif Walsh, Michael
  Bender, Martin Farach-Colton, Rob Johnson, Bradley~C. Kuszmaul, and Donald~E.
  Porter.
\newblock Betr{FS}: A right-optimized write-optimized file system.
\newblock In {\em 13th USENIX Conference on File and Storage Technologies (FAST
  15)}, pages 301--315, Santa Clara, CA, USA, 2015. USENIX Association.

\bibitem{JannenYuZh15tos}
William Jannen, Jun Yuan, Yang Zhan, Amogh Akshintala, John Esmet, Yizheng
  Jiao, Ankur Mittal, Prashant Pandey, Phaneendra Reddy, Leif Walsh, Michael
  Bender, Martin Farach-Colton, Rob Johnson, Bradley~C. Kuszmaul, and Donald~E.
  Porter.
\newblock {BetrFS}: Write-optimization in a kernel file system.
\newblock {\em ACM Transactions on Storage}, 11(4), November 2015.

\bibitem{JiChSh16}
Cheng Ji, Li-Pin Chang, Liang Shi, Chao Wu, Qiao Li, and Chun~Jason Xue.
\newblock An empirical study of file-system fragmentation in mobile storage
  systems.
\newblock In {\em 8th USENIX Workshop on Hot Topics in Storage and File Systems
  (HotStorage 16)}, Denver, CO, 2016. USENIX Association.

\bibitem{DBLP:conf/hotstorage/JiCSWLX16}
Cheng Ji, Li{-}Pin Chang, Liang Shi, Chao Wu, Qiao Li, and Chun~Jason Xue.
\newblock An empirical study of file-system fragmentation in mobile storage
  systems.
\newblock In Nitin Agrawal and Sam~H. Noh, editors, {\em HotStorage}. {USENIX}
  Association, 2016.

\bibitem{JuKa13}
Myoungsoo Jung and Mahmut Kandemir.
\newblock Revisiting widely held ssd expectations and rethinking system-level
  implications.
\newblock In {\em Proceedings of the ACM SIGMETRICS/international conference on
  Measurement and modeling of computer systems (SIGMETRICS '13)}, pages
  203--216, New York, NY, USA, 2013. ACM.

\bibitem{kadekodi18atc}
Saurabh Kadekodi, Vaishnavh Nagarajan, and Gregory~R. Ganger.
\newblock Geriatrix: Aging what you see and what you {don{\textquoteright}t}
  see. a file system aging approach for modern storage systems.
\newblock In {\em USENIX ATC}, pages 691--704, Boston, MA, July 2018. USENIX
  Association.

\bibitem{KarlinMaMc94}
Anna~R. Karlin, Mark~S. Manasse, L.~A. McGeoch, and S.~Owicki.
\newblock Competitive randomized algorithms for nonuniform problems.
\newblock {\em Algorithmica}, 11(6):542--571, 1994.

\bibitem{f2fstools}
Jaegeuk Kim.
\newblock f2fstools: Userland tools for the f2fs filesystem.
\newblock
  https://git.kernel.org/pub/scm/linux/kernel/git/jaegeuk/f2fs-tools.git.
\newblock Version 1.15.0. Accessed 17 February 2023.

\bibitem{lee15f2fs}
Changman Lee, Dongho Sim, Jooyoung Hwang, and Sangyeun Cho.
\newblock {F2FS}: A new file system for flash storage.
\newblock In {\em 13th USENIX Conference on File and Storage Technologies (FAST
  ’15)}, pages 273--286, Santa Clara, CA, USA, February~22--25 2015.

\bibitem{MaFeLi14}
Dongzhe Ma, Jianhua Feng, and Guoliang Li.
\newblock A survey of address translation technologies for flash memories.
\newblock {\em ACM Comput. Surv.}, 46(3):36:1--36:39, January 2014.

\bibitem{MathurCaBh07}
Avantika Mathur, MingMing Cao, Suparna Bhattacharya, Andreas Dilger, Alex
  Tomas, and Laurent Vivier.
\newblock The new ext4 filesystem: current status and future plans.
\newblock In {\em Ottowa Linux Symposium (OLS)}, volume~2, pages 21--34,
  Ottowa, ON, Canada, 2007.

\bibitem{McKusickJoLe84}
Marshall~K. McKusick, William~N. Joy, Samuel~J. Leffler, and Robert~S. Fabry.
\newblock A fast file system for {UNIX}.
\newblock {\em ACM Transactions on Computer Systems}, 2(3):181--197, August
  1984.

\bibitem{MinKiCh12}
Changwoo Min, Kangnyeon Kim, Hyunjin Cho, Sang{-}Won Lee, and Young~Ik Eom.
\newblock {SFS:} random write considered harmful in solid state drives.
\newblock In {\em 10th USENIX Conference on File and Storage Technologies (FAST
  12)}, pages 139--154, San Jose, CA, USA, February~14--17 2012.

\bibitem{OneilChGa96}
Patrick O'Neil, Edward Cheng, Dieter Gawlic, and Elizabeth O'Neil.
\newblock The log-structured merge-tree ({LSM}-tree).
\newblock {\em Acta Informatica}, 33(4):351--385, 1996.

\bibitem{oracleXFSdefrag}
Oracle.
\newblock Defragmenting an {XFS} file system.
\newblock Oracle Linux Administrator's Solution Guide for Release 6.
\newblock Accessed 10 May 2016.

\bibitem{park21sosp}
Jonggyu Park and Young~Ik Eom.
\newblock {\em FragPicker: A New Defragmentation Tool for Modern Storage
  Devices}, pages 280--294.
\newblock ACM, New York, NY, USA, 2021.

\bibitem{ParkKaEo16}
Jonggyu Park, Dong~Hyun Kang, and Young~Ik Eom.
\newblock File defragmentation scheme for a log-structured file system.
\newblock In {\em Proceedings of the 7th ACM SIGOPS Asia-Pacific Workshop on
  Systems}, APSys '16, pages 19:1--19:7, New York, NY, USA, 2016. ACM.

\bibitem{RodehBaMa13}
Ohad Rodeh, Josef Bacik, and Chris Mason.
\newblock {BTRFS}: The {L}inux {B}-tree filesystem.
\newblock {\em ACM Transactions on Storage}, 9(3), August 2013.

\bibitem{RoselliLoAn01}
Drew Roselli, Jacob~R. Lorch, and Thomas~E. Anderson.
\newblock A comparison of file system workloads.
\newblock In {\em Proceedings of the Annual Conference on USENIX Annual
  Technical Conference}, ATEC '00, pages 4--4, Berkeley, CA, USA, 2000. USENIX
  Association.

\bibitem{RosenblumOu92}
Mendel Rosenblum and John~K. Ousterhout.
\newblock The design and implementation of a log-structured file system.
\newblock {\em ACM Transactions on Computer Systems (TOCS)}, 10(1):26--52,
  February 1992.

\bibitem{SmithSe97}
Keith~A. Smith and Margo Seltzer.
\newblock File system aging --- increasing the relevance of file system
  benchmarks.
\newblock In {\em Proceedings of the 1997 ACM SIGMETRICS international
  conference on Measurement and modeling of computer systems (SIGMETRICS '97)},
  pages 203--213, Seattle, WA, June~15--18 1997.

\bibitem{SweeneyDoHu96}
Adam Sweeney, Doug Doucette, Wei Hu, Curtis Anderson, Mike Nishimoto, and Geoff
  Peck.
\newblock Scalability in the {XFS} file system.
\newblock In {\em Proceedings of the USENIX 1996 Annual Technical Conference},
  San Diego, CA, USA, January22--26 1996.

\bibitem{e2freefrag}
Rupesh Thakare, Andreas Dilger, and Kalpak Shah.
\newblock e2freefrag(8) - linux man page.
\newblock \url{https://linux.die.net/man/8/e2freefrag}.

\bibitem{Tso15}
Theodore Ts'o.
\newblock E2fsprogs: Ext2/3/4 filesystem utilities, May~17 2015.
\newblock Version 1.42.13. Accessed 10 May 2016.

\bibitem{Tweedie00}
Stephen Tweedie.
\newblock {EXT3}, journaling filesystem.
\newblock In {\em Ottowa Linux Symposium}, Ottowa, ON, Canada, July~20 2000.

\bibitem{WirzeniusOjSt04}
Lars Wirzenius, Joanna Oja, Stephen Stafford, and Alex Weeks.
\newblock {\em Linux System Administrator's Guide}.
\newblock The Linux Documentation Project, 2004.
\newblock Version 0.9.

\bibitem{YuanZhJa16}
Jun Yuan, Yang Zhan, William Jannen, Prashant Pandey, Amogh Akshintala, Kanchan
  Chandnani, Pooja Deo, Zardosht Kasheff, Leif Walsh, Michael Bender, Martin
  Farach-Colton, Rob Johnson, Bradley~C. Kuszmaul, and Donald~E. Porter.
\newblock Optimizing every operation in a write-optimized file system.
\newblock In {\em 14th USENIX Conference on File and Storage Technologies (FAST
  16)}, pages 1--14, Santa Clara, CA, USA, 2016.

\bibitem{yuan17tos}
Jun Yuan, Yang Zhan, William Jannen, Prashant Pandey, Amogh Akshintala, Kanchan
  Chandnani, Pooja Deo, Zardosht Kasheff, Leif Walsh, Michael~A. Bender, Martin
  Farach-Colton, Rob Johnson, Bradley~C. Kuszmaul, and Donald~E. Porter.
\newblock Writes wrought right, and other adventures in file system
  optimization.
\newblock {\em ACM Trans. Storage}, 13(1), mar 2017.

\bibitem{zhan18fast}
Yang Zhan, Alex Conway, Yizheng Jiao, Eric Knorr, Michael~A. Bender, Martin
  Farach-Colton, William Jannen, Rob Johnson, Donald~E. Porter, and Jun Yuan.
\newblock The full path to {Full-Path} indexing.
\newblock In {\em 16th USENIX Conference on File and Storage Technologies (FAST
  18)}, pages 123--138, Oakland, CA, February 2018. USENIX Association.

\bibitem{zhan21tos}
Yang Zhan, Alex Conway, Yizheng Jiao, Nirjhar Mukherjee, Ian Groombridge,
  Michael~A. Bender, Martin Farach-Colton, William Jannen, Rob Johnson,
  Donald~E. Porter, and Jun Yuan.
\newblock Copy-on-abundant-write for nimble file system clones.
\newblock {\em ACM Trans. Storage}, 17(1), jan 2021.

\bibitem{zhan20fast}
Yang Zhan, Alexander Conway, Yizheng Jiao, Nirjhar Mukherjee, Ian Groombridge,
  Michael~A. Bender, Martin Farach-Colton, William Jannen, Rob Johnson,
  Donald~E. Porter, and Jun Yuan.
\newblock How to copy files.
\newblock In {\em 18th USENIX Conference on File and Storage Technologies (FAST
  20)}, pages 75--89, Santa Clara, CA, February 2020. USENIX Association.

\bibitem{zhan18tos}
Yang Zhan, Yizheng Jiao, Donald~E. Porter, Alex Conway, Eric Knorr, Martin
  Farach-Colton, Michael~A. Bender, Jun Yuan, William Jannen, and Rob Johnson.
\newblock Efficient directory mutations in a full-path-indexed file system.
\newblock {\em ACM Trans. Storage}, 14(3), nov 2018.

\bibitem{ZhuChCh05}
Ningning Zhu, Jiawu Chen, and Tzi-Cker Chiueh.
\newblock {TBBT}: Scalable and accurate trace replay for file server
  evaluation.
\newblock In {\em 4th USENIX Conference on File and Storage Technologies (FAST
  05)}, pages 323--336, Santa Clara, CA, USA, 2005.

\end{thebibliography}

\end{document}